\documentclass[review]{elsarticle}

\pdfoutput=1

\usepackage{hyperref}
\usepackage[disable]{todonotes}
\usepackage{amsmath,amssymb}
\usepackage{textcomp}
\usepackage{fancyhdr}
\usepackage{multicol}
\usepackage{color}
\usepackage{hyphenat}
\usepackage[scaled=1]{helvet}
\usepackage[normalem]{ulem}
\usepackage[utf8]{inputenc}
\usepackage{caption}

\usepackage{url}

\journal{Journal of Petroleum Science and Engineering}

\usepackage{numcompress}
\biboptions{authoryear}

\newcommand{\TODO}[1]{}

\begin{document}

\begin{frontmatter}

\title{A Decision Support System for Multi-target Geosteering
}

\author[norce]{Sergey Alyaev\corref{correspond}}
\author[norce]{Erich Suter}
\author[uis]{Reider Brumer Bratvold}
\author[uis]{Aojie Hong}
\author[norce]{Xiaodong Luo}
\author[norce]{Kristian Fossum}

\address[norce]{NORCE Norwegian Research Centre, Postboks 22 Nygårdstangen, 5838 Bergen, Norway}
\address[uis]{University of Stavanger, Postboks 8600 Forus, 4036 Stavanger, Norway}

\cortext[correspond]{Corresponding author: Sergey.Alyaev@norceresearch.no}

\begin{abstract}
Geosteering is a sequential decision process under  uncertainty. 
The goal of geosteering is to maximize the expected value of the well, which should be defined by an objective value-function for each operation. 

In this paper we  present a real-time decision support system (DSS) for geosteering that aims to approximate the uncertainty in the geological interpretation with an ensemble of geomodel realizations. 
As the drilling operation progresses, the ensemble Kalman filter is used to sequentially update the realizations using the measurements  from real-time logging while drilling. 
At every decision point 
a discrete
dynamic programming algorithm computes 
all potential well trajectories for the entire drilling operation 
and
the corresponding value of the well for each realization. 
Then, the DSS considers all immediate alternatives (continue/steer/stop) and chooses the one that gives the best predicted value across the realizations. 
This approach works for a variety of objectives and constraints and suggests
reproducible decisions under uncertainty. Moreover, it has real-time performance.

The system is tested on synthetic cases in a layer-cake geological environment where the target layer should be selected dynamically based on the prior (pre-drill) model and the electromagnetic observations received while drilling. 
The numerical  closed-loop simulation experiments 
demonstrate the ability of the DSS to perform successful geosteering and 
landing of a well for different geological configurations of drilling targets. 
Furthermore, the DSS allows to adjust and re-weight the objectives, 
making the DSS useful before fully-automated geosteering becomes reality.
 \end{abstract}


\end{frontmatter}

%\linenumbers

\section{Introduction}\label{secIntroduction}

According to the Norwegian Petroleum Directorate , drilling new wells is the most efficient way to increase oil recovery \citep{npd2018}. 
At the same time, well delivery and maintenance constitutes one of the major costs of oil reservoir development \citep{saputelli2003real}.
To maximize value creation from each well, 
operators and service companies are continuously improving technology and procedures for 
optimizing the well placement to maximize production while reducing the cost of drilling and future maintenance. 
To place a well precisely in the best reservoir zone, operators use geosteering to adjust the well trajectory 
in response to 
real-time information acquired while drilling.
The benefits of geosteering, such as higher production rates of the resulting wells, have been 
extensively documented in the literature 
\citep{al2004increased,janwadkar2012reservoir,guevara2012milestone, ENI2017GoliatCaseStudies}.

\todo[inline]{FIXED Rev1: Page 2 Line 14: A minor comment on the terms 'manual geological interpretation'. As you might already know, computers have long been used to support interpretation. While I understand the context of this sentence, the terms could cause some readers to disregard the work as misunderstanding the industry practice.}

Geosteering has traditionally been dominated by manual geological interpretation and decision-making. 
Current computer-aided approaches assist decision-makers by co-visualizing a pre-drill deterministic geomodel alongside the inversion results of real-time data deep resistivity data. 
It is then 
up to the team of geoscientists 
to interpret the available information and decide steering actions in real-time (see e.g. \cite{Boe2014}). 

More recently, there has been a focus on advancing computer-based methods both for 
pre-job, 
post-job and 
real-time analysis to support interpretation during drilling.
The paper \citet{Antonsen2018Essential} discusses the importance of establishing a good understanding of how essential reservoir objects such as top reservoir and oil-water contact (OWC) are mapped
by inversion of deep electromagnetic measurements in the pre-job phase of the drilling operation. 
This concept is extended in \citet{Equinor2018AntonsenIntegration}, where post-job case studies illustrate the importance of combining multiple LWD measurements with pre-job geophysical modeling. 
\cite{SaudiAramco2016ReservoirModellingGeosteering} have explained how the formation tops in a 3D geo-cellular near-well sector model were adjusted in depth during drilling. 
In \cite{ENI2017ArataIntegrationLWDSeismic} a case study shows how a real-time
local recalibration of seismic to minimize the depth discrepancy, based on LWD measurements, supports improved prediction of the 
reservoir boundaries ahead of the bit.
Finally, in \citep{payrazyan2017geoscience} it is explained how the geological structures (faults and stratigraphic interfaces) in a 2D section along the well can be adjusted in real-time to fit the measurements, as basis for real-time sketching of a desired trajectory in the graphical interface.

The workflows discussed above are steps towards automated inversion and
interpretation of real-time measurements.
However, 
the information extracted from data has no value unless it helps us make better decisions. 
Geosteering is fundamentally about making 
decisions to optimize outcomes such as achieving optimal production at minimal costs.
Making decisions that honor all different and sometimes conflicting objectives is not intuitive and requires excessive calculations that can only be handled by a computer.

Currently there is a lack of methods, tools and workflows
that explicitly treat the uncertain nature of this decision process.
To optimize the well placement under uncertainty, we should work within a probabilistic framework 
using a dedicated decision-analytic framework \citep{Kullawan2014}.
The first step is utilization of prior data and descriptive analytics 
to summarize and improve our probabilistic understanding 
of the reservoir formation. 
Thereafter the real-time measurements provide
information that improves our understanding of the geological and operational parameters that are 
crucial to optimal well placement. 
Finally, predictive analytics supports the continuous updating of our 
understanding of these parameters, and gives input to decisions on
directional changes or stopping.

\todo[inline]{Rev1:
Page 4 Line 67: The real-time update of the realizations should reduce both the interpretation uncertainty (human limitations to interpret the data) and measurement uncertainty (higher resolution of real-time measurement compared to pre-drilled measurement).}
\todo[inline]{@Sergey: Disagree.}
\todo[inline]{@ECS: the comment looks OK to me. The first part is definitely OK, and the second part probably refers to that uncertainties in the seismic measurements can be reduced when considered in light of high-resolution LWD measurements (simply stated), which is a form of interpretation (you can believe more in the seismic if confirmed by high resolution measurements). My guess is that this comment is a response to the next sentence, which I commented on below.}
\todo[inline]{@ECS2: I cannot see that this this comment has been addressed in the text ...}
\todo[inline]{@Sergey2: I read and the issue does not exist. Was a minor issue any way}

In this paper we present a consistent, systematic and transparent workflow for geosteering, which implements the principles above in a computer-based decision support system (DSS).
The starting point is 
a probabilistic geomodel 
represented by multiple geomodel realizations which
aim to span the space of pre-drill interpretation 
uncertainties.
The real-time measurements obtained while drilling are continuously integrated 
by automatically updating the realizations using an ensemble-based filtering method, similar 
to \cite{Chen2015spe,Luo2015}.
The real-time update of the realizations aims to provide an always up-to-date prediction of the subsurface including interpretation uncertainty.
\todo{Unclear, and has no value here: The update workflow is linked to the DSS.}

The update workflow
is linked to the decision optimization.
The DSS uses the up-to-date probabilistic  geomodel
to support geosteering decisions under uncertainty.
It proposes well trajectories ahead of the bit and evaluates them  against the chosen value function.
\todo{[was] Unclear}
The value function commonly includes multiple objectives, including production potential, costs for 
drilling and completion, and risks associated with the operation.
The evaluation of trajectories is basis for the optimization.
The trajectory optimization in the DSS is inspired by the discretized stochastic dynamic programming 
algorithms for geosteering that were discussed in
\cite{kullawan2017decision,kullawan2018}. 
However, the DSS presented here is specifically 
optimized for usage with ensemble-based update workflows 
which are already used for field development planning \citep{hanea2015reservoir,skjervheim2015fast}.

The real-time update workflow was previously demonstrated for pro-active geosteering with the objective to follow the top of a reservoir \citep{Chen2015spe,Luo2015}. 
The DSS presented in this paper combines this update workflow with dynamic programming for global trajectory optimization under uncertainty. The new optimization algorithm enables a variety of practical objectives,  which among other things allow to optimize well-landing in uncertain environments. 

The goal of this paper is to verify the DSS workflow on comprehensive synthetic experiments.
Our numerical experiments are inspired by a challenging set-up with multiple target layers from a case study presented in \cite{Hongsheng2016}.
Unlike an expert service required for successful geosteering in the mentioned case, the DSS delivers reproducible and good decisions under uncertainty which maximize the set of objectives selected for an operation.
We presented the flexibility of the DSS with respect to selection of objectives and initial tests in earlier conference proceedings \citep{Alyaev2018,Alyaev2018d}. 
Here we focus on exhaustive presentation of the features and limitations of the DSS algorithms and present a statistical verification of the performance of the system.

The rest of the paper is organized as follows. Firstly, we present the ensemble-based workflow for 
updating of the probabilistic geomodel based on real-time measurements. 
Secondly, we introduce the DSS that 
can utilize the up-to-date ensemble to propose optimal decisions under uncertainty.
After that, the performance of the DSS is
demonstrated on synthetic cases with multiple targets. 
Finally the main contributions of the paper are summarized in
the conclusions.
 
\section{Earth model update loop}

\begin{figure}
  \centering
  \includegraphics[width=0.8\textwidth]{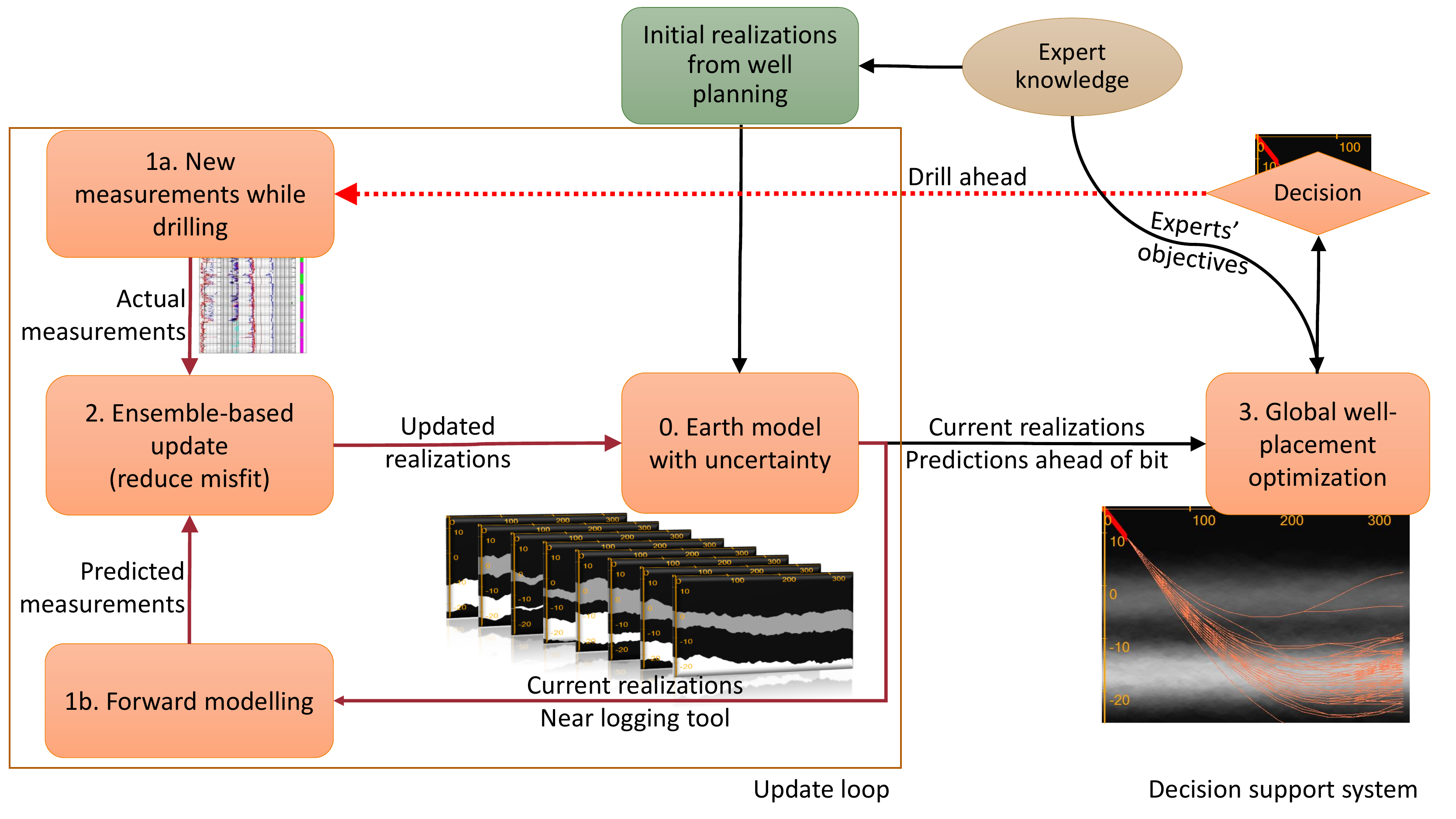}
  \caption{Proposed geosteering workflow. 
  The top part contains inputs to the workflow.
  The left part of the figure depicts the update loop.
  The part of the figure to the right contains the decision system that is based on the updated 
earth 
model.
  The 'drill ahead' decision results in new measurements that trigger another update and 
complete the 
full loop of the workflow.}
  \label{figWorkflow}
\end{figure}

In the proposed geosteering workflow, shown in Figure \ref{figWorkflow}, real-time decision support 
is based on Bayesian inference from
a probabilistic geomodel that is continuously updated.
\subsection{Earth model}
The earth model is represented as an ensemble of realizations that captures key
geological uncertainties. In figure \ref{figWorkflow} the uncertainties are the positions and thicknesses of sand layers (gray) in a background shale.
The pre-drill realizations are created based on a priori information drawn from seismic, logs from offset wells, production 
measurements 
and additional knowledge about geological uncertainties provided by experts.  

All realizations are updated incrementally each time new measurements become available while drilling.
The incremental updates of the model are performed by a 
statistically-sound ensemble-based method that reduces the mismatch between the measurements and 
the geomodel. 
The ensemble-based updating approach is an implementation of a Bayesian updating framework. 
In the rest of the section we describe the implementation of the individual components 
of the generic update loop that was used in this study.

\subsection{Measurements}
\noindent
By design, the ensemble-based methods perform incremental updates which can handle any number and 
any type of 
measurements  simultaneously (see 1a. in figure 
\ref{figWorkflow}).
It is required however, that there is a corresponding 
simulation model that can transform the realizations and the measurements to a context where they 
can be adequately compared to compute the mismatch. The simplest way is to use a forward model that 
produces synthetic measurements based on the geomodel realizations (see 1b. in figure 
\ref{figWorkflow}). The forward model should be 
sufficiently fast to handle hundreds of simulations at every assimilation step. 

\begin{figure}
\centering
 \includegraphics[width=0.5\textwidth,trim={0 200px 259px 0},clip]{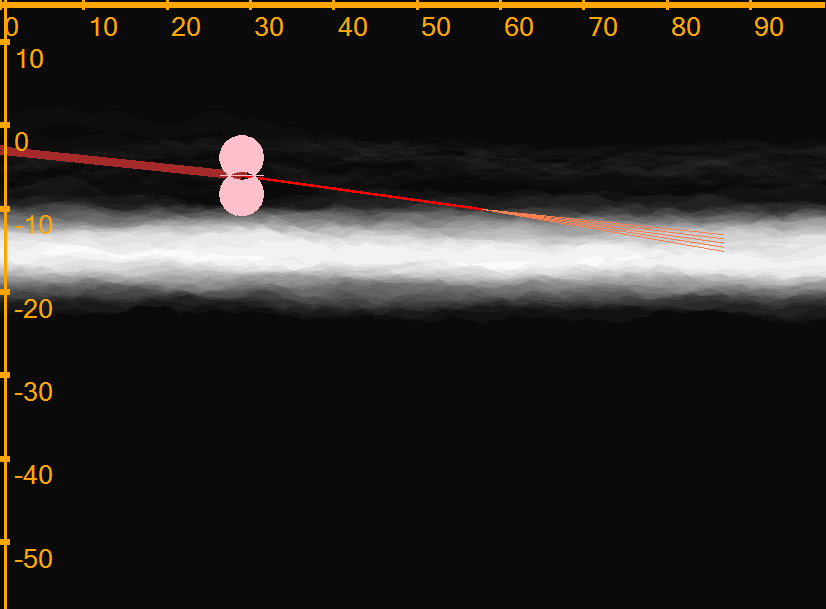}
 \caption{An illustration of the depth of investigation (DOI) of the tool for the synthetic model (axes 
in meters). The dark red line segment to the left shows the trajectory that has been drilled; the 
thin red line segment 
shows the next proposed trajectory segment. The DOI is illustrated at the 
current decision point at the end of the drilled trajectory. The measurements at the highlighted 
location have been assimilated. 
}
 \label{figDepthInvestigation}
\end{figure}

\TODO{Regarding: ``The DOI is illustrated at the current decision point at the end of the drilled 
trajectory''.
The DOI is placed around the BHA, and should be 10-20 m behind the bit. But this is only a synthetic model,
so it doesn't matter. But it should be commented in the text that the DOI in reality will be placed somewhat behind
the bit.}

\TODO{Comment: ``One can notice that the uncertainty of the layer boundary near the 
tool has been reduced compared to the uncertainty of the bottom horizon.'' This is not clear in the figure, so I removed it.}

\subsection{Forward modelling}
\noindent
The simulation methods for processing of different logs have been extensively studied by service 
companies 
\citep{Sviridov2014a,Dupuis2014,Dupuis2015,Hartmann2014a,selheim2017geosteering}
but are generally not available in the open domain, only as paid services.
The main contribution of this paper is the DSS and not the modelling of 
the measurements. Therefore, we use a simple integral model for electromagnetic (EM) measurements 
following
\cite{Chen2015spe}. 
The tool set-up in that paper has a look-around capability of about 5 meters, and 
it is sensitive to resistivity
in the up, down and side directions, see figure 
\ref{figDepthInvestigation}.
The 
tool is placed at the drill-bit in the current prototype.
The depth of investigation (DOI) is chosen relatively low compared to the modern deep EM tools (e.g. \citet{Seydoux2014}) to maintain 
the accuracy of the 
approximate integral model.
However, we emphasize that this does not constrain the applicability of the workflow. 
For instance, in \cite{Luo2015}, a similar update 
workflow has been tested with more advanced tools and a finite difference forward model. The tool 
modeled in \cite{Luo2015} provides a higher DOI that allows to see in a
larger volume
around the well and is expected to yield better results.

\subsection{Ensemble-based update algorithm}
\noindent
The update loop used in this paper is  compatible with a number of ensemble-based methods
which have previously been implemented for reservoir 
data assimilation including the ensemble Kalman filter \citep{aanonsen2009ensemble},
ensemble smoother \citep{skjervheim2011ensemble,skjervheim2015fast},
the particle filter \citep{lorentzen2016estimation},
and more sophisticated combinations of the above, such as adaptive Gaussian mixture filter 
\citep{lorentzen2017auxiliary}.
To demonstrate the workflow, we use the standard ensemble Kalman filter (EnKF) method 
\citep{Chen2015spe,Luo2015} for the 
implementation 
described in this paper.

The Kalman  filter \citep{kalman1960new} formulates the Bayesian update for the changes of the mean and the covariance matrix when new data is received assuming all probability distributions are Gaussian. 
The ensemble Kalman filter \citep{evensen1994sequential,evensen2009data} is a 
flexible Monte Carlo implementation of the Kalman filter.
When new measurements are received, they are compared to the simulated measurements generated by 
the 
corresponding forward models for each realization.
The realizations used in this paper assume a layer-cake geomodel with constant 
resistivity in each layer. 
The depth of each layer boundary  is represented by a series of points. 
\todo{SEAL: looks good... Next sentence: should be '... yield an update...'. Each update is not incremental, but you get incremental updates as the drilling progresses.}
The new measurements
yield an 
incremental update of the depth values in the interfaces, which can be formulated as in \citet{burgers1998analysis}:
\begin{equation}
	y_{updated} = y_{initial} + K (d_{measured} - d_{modelled}),
	\label{eqEnkfUpdate}
\end{equation}
where $y_{updated}$ contains the updated (or posterior) ensemble representing the posterior distribution, 
$y_{initial}$ contains the initial (or prior) ensemble representing the prior distribution, $K$ is the Kalman gain matrix, $d_{measured}$ contains the perturbed measured data values\footnote{
For EnKF, a measured data value has to be perturbed with its corresponding statistics in order to avoid insufficient variance \citep{burgers1998analysis}.
},
and $d_{modelled}$ contains the modelled data values corresponding to the initial ensemble for that update. 
Equation \eqref{eqEnkfUpdate} describes a linear combination of the prior and the measured data, which is weighted by the 
Kalman gain matrix $K$. 
Bayes' rule describing the relationship among the prior, likelihood, and posterior is not shown explicitly in the equation above, but it is implicitly included as the likelihood is encoded in the Kalman gain matrix $K$ and the pre-posterior is treated as a normalizing constant of the updated ensemble. 
 A detailed description on the relationship between the formulation of the Kalman filter and the Bayesian formulation can be found in \cite{meinhold1983understanding}.

The incremental nature of the updates as drilling progresses removes the need for 
a costly direct inversion that include all the available measurements 
each time the model is updated.
By design  the updates (e.g. depth of boundaries) are also
propagated ahead of the bit using the prior knowledge about the model.
\todo{Prev sentence: this is not entirely true, updates e.g. in the existence of faults or layers is not updated. Experts within the field would pick on this (it is an important and well known limitation which has been discussed in geomodelling papers).
Could write 'By design the updates in the boundary depths are also propagated ...'}
This  provides a probabilistic prediction 
of the geology ahead of the bit based on the 
trends identified around and behind the bit.
We refer the reader to \cite{Luo2015} for a more rigorous description of the update loop for geosteering.
 
\section{Decision support system (DSS)}

The update loop described above results in an always up-to-date 
ensemble of model realizations
which integrates both the prior knowledge and the latest measurements.
The realizations representing the probabilistic description of relevant and material geological uncertainties are
the input to the DSS. 
The DSS is based on a normative decision-making approach \citep{bratvold2010making,clemen2013making,howard2015foundations} and includes optimization algorithms that take into account all realizations as well as 
multiple decision
objectives, such as following the reservoir top while minimizing drilling cost and reducing the 
tortuosity of the well for easier completion.
The objectives may be conflicting, and for each realization the 
algorithm calculates the  well path
that is optimal with respect to the weight of each objective. 
The objectives and the 
corresponding weights are defined by the user of the DSS and is a consistent way to 
include expert knowledge in the workflow, see figure \ref{figWorkflow}.
The proposed decision for stopping or adjusting the trajectory is visualized together with the 
current 
representation of the uncertainty in the model. 

The DSS presented in this paper differs from traditional decision systems that were 
designed for strategic decisions.
In contrast, the geosteering decisions are operational. 
This implies that they must be taken within short time.
The presented DSS allows to tweak the weights of the objectives at all decision points and preview the outcomes in real-time.
This helps 
the user to build an understanding of how the choice of objectives influence the suggested 
decisions and provides a possibility to re-evaluate the trade-offs between objectives as the 
drilling progresses.

The logically  consistent approach of the DSS allows for decisions to be transparent and reproducible. 
Given that the DSS is based on a normative decision quality approach, it will recommend good decisions for the decision-maker's objectives, alternative choices for a decision (following constraints), and geological beliefs (represented in the pre-drill model).
\TODO{Aojie : I think this statement is used exclusively for the farsighted approach. For the naive-optimistic approach used in the DSS, the assumption of perfect information is inconsistent with our beliefs. However, we use this assumption for simplification and computationally feasibility. We don't need to spell this out in this paper, but it is for our own clarity.}
\todo{Prev sentence: 'alternatives (following constraints)' what is meant by this?}

\subsection{Objectives}

A natural requirement for any DSS is the possibility to take into account multiple 
objectives. The objectives used in modern geosteering operations include placing the well in 
a specific position in the reservoir, reducing costs and ensuring safety \citep{Kullawan2016value}.
For the use in a DSS the objectives need to be converted into objective functions defined on a common scale,
e.g. the estimated profit in US dollars or produced-oil-barrels equivalent. 
To reduce conversions, we will use the stand length 
drilled within standard reservoir sand as the common scale in this paper. 
We denote each objective 
function as 
$O_i(X \vert  M)$, which depends on  the trajectory ($X$) of the well  and the actual sub-surface configuration ($M$), which for now we assume to be known and deterministic. 
The profit functions are positive and costs 
associated with the operation are negative. 
Objective functions that are used in our numerical examples are summarised in the appendix (Section \ref{secAppendix}).
\todo[inline]{fixed?: 
Page 11 Line 221-224: My first thought after reading equation (2) is $w_i$ represent relative weight among objectives. I, then, intuitively assume  $\sum w_i = 1$. However, from line 221-224, it was clear that I misunderstood. It could be a good idea to explain more on this and also how we arrive at  $w_i$ on each objective.}

The global objective $O(X \vert  M)$ is represented as a linearly weighted sum of individual 
contributions from each objective function:
\begin{equation}
  O(X \vert  M) 
  = 
  \sum_{i}
  w_i O_i(X \vert  M),
  \label{eqGlobalObjective}
\end{equation}
where $0 \leq w_i$ is an objective weighting factor for objective $i$, where $i$ are indices of 
different objectives. 
\todo{Prev sentence: 'objective weighting factor': the weighting factor is subjective as far as I know (set by the user, based on a qualitative belief). I also see from the TODO on line 94 that Reidar comments on this, so maybe better to use his explanation of why the weighting factor should be objective (and how to achieve this).}
The functions $O_i$ are scaled so that the initially estimated (pre-drill)
profit/cost corresponding to each objective function is achieved when $w_i = 1$. 
It is convenient to think of $w_i = 0$ as ignoring the objective $i$, while $w_i = 1$ means setting the value of objective $i$ to the scale anticipated in the pre-drill analysis.
\TODO{Reidar: In a consistent probabilistic framework, there should be no initial base-case type estimate as this tends to lead to a set of biases. Rather, the initial estimate should be a distribution or, more commonly, a set of percentiles (say 10/50/90-percentile) which represent the DMs beliefs.}
\todo{Changed default to pre-drill analisys}
Furthermore, adjusting the weights gives the user
the control 
to modify the priority of the objectives
and
to maintain the predictions at a desired scale, 
both in response to insights gained during the drilling operation.\footnote{In most situations $\sum_{i}  w_i \neq 1$.}
\todo{[corrected] Prev sentence: As I see it, 'insights' belong to priority of the objectives, not to the desired value scale. But I'm not the expert here... Can the value scale change while drilling?}
$w_i > 1$ corresponds to a
higher priority to objective $O_i$, while $w_i < 1$ corresponds to a lower priority. 
Changing the weights reflects an insight
in how each objective contributes to the profit/cost of the well being drilled compared to the
originally anticipated (see the last numerical example in the next section).
We will use the global objective defined by \eqref{eqGlobalObjective} as the value function in the 
optimization for the rest of the section.

\subsection{Sequential decision optimization under uncertainty}
\label{secSecventialDecision}
\TODO{I do not like the position of this sub-section}
A geosteering operation consists of a sequence of decisions $D_k$ .
Subscript $k$ numerates decision points sequentially in time.
Ensemble-based workflows represent 
the uncertainty in the geological interpretation as a set of 
realizations. 
Substituting different realizations $M_j$ instead of the deterministic model $M$ into the objective function in equation \eqref{eqGlobalObjective} typically gives several trajectories, where each is optimal for the corresponding realizations.
At the same time, the outcome of the optimization should be a single optimal decision for each decision point $k$. 
In this paper we follow the optimality criterion used in robust optimization:
We want to make a decision $D_k$  that maximizes the expected value of the well given all the available information.

\begin{figure}
	\centering
	\includegraphics[width=\textwidth]{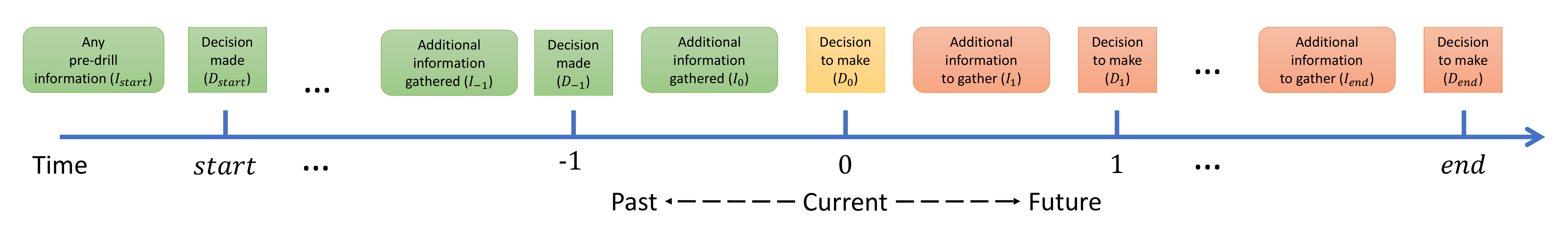}
	\caption{Full structure of a sequential decision problem}
	\label{figDecisionProblem}
\end{figure}

Let us consider all available information at time $k$.
At each decision point the ensemble-based workflow contains up-to-date realizations representing the current understanding of the subsurface.
Moreover, between any two sequential decision points, 
new measurements are assimilated using the update loop. This improves the geological understanding around and ahead of the drill-bit.
The full structure of a sequential  decision problem is shown
in Figure \ref{figDecisionProblem}, 
where $D_k$ denotes the decision at time $k$, and $I_k$ denotes the information gathered between time $k-1$ and $k$. 
For brevity of notation we denote all information gathered between $k_1$ and $k_2$ as  $I_{(k_1:k_2)}$.
Generally, at some current time $k=0$, the decision $D_0$,
that needs to be made right now, depends not only on the information that has been gathered $I_{(start:0)}$ and is contained in the geomodel,
 but also on the possibility of future learning, i.e. the
information that will be gathered $I_{(1:end)}$.
Because the future information $I_{(1:end)}$ is not available at time $k$, its influence
on future learning and decision-making
 can be modelled 
as uncertain events,
conditioned 
on the current information and prior decisions.

The approach considering the full learning and decision-making structure of a sequential 
decision-making problem is presented in Figure \ref{figDecisionProblem}.
We call this approach "far-sighted"  as it takes into account what might happen in the future 
including which 
information that
will be gathered, how uncertainties will be updated using that information, and which decisions that will be made \citep{Alyaev2018b}.
An implementation of the far-sighted approach using discretized stochastic dynamic programming has been described in detail in \cite{kullawan2017decision,kullawan2018} for a geosteering problem considering a geomodel with a single reservoir layer and updates of its boundaries.

Unfortunately, the stochastic modelling required for understanding the effect of future learning in the far-sighted approach is computationally prohibitive for real-time decision-making.
First, the complexity of the problem grows exponentially as the numbers of decision points, alternatives and uncertainty branches increase. 
\todo{looks OK Prev sentence: which alternatives? realizations? objectives?}
The phenomenon is known as the curse of dimensionality, see \cite{Brown2013}. Thus, the far-sighted approach becomes computationally prohibitive for problems with a  large number of parameters.
Second, the state-of-the-art methods for data assimilation (e.g., the ensemble Kalman filter used in our update loop) cannot be directly embedded into the far-sighted approach. 
The far-sighted approach requires generating and storing not only realizations for the current decision point, but also realizations that are modified due to future updates in the EnKF loop, for all future decision points.

\TODO{
In the DSS framework, the aim is to handle any type of objective function provided by
the geosteering decision team. 
Hence, it is necessary to apply a global optimization algorithm. We define a global 
optimization algorithm as one that 
optimizes the complete well path ahead of the bit against the currently available 
representation of the geological uncertainty (as represented by the set of realizations) and
does not fall into local minima.}

In this paper we present a new dynamic programming discrete optimization strategy 
which has real-time performance and
is simple to integrate with the ensemble-based update loop.
The strategy is a simplification of the far-sighted approach.
It 
considers future decisions but omits the modelling required to simulate the future learning.
Instead of the modelling of the future information we optimistically assume
that perfect information (about the subsurface)
would be available after the current decision has been made and before the next decision is made.
\todo{Prev sentence: what does 'information' mean,  e.g. '... information, in this case ...'.}
Thus, this approach can be classified as naive optimistic \citep{Alyaev2018b}. 
\TODO{Assuming the perfect information means that the expected value is overestimated}
It is possible to find theoretical scenarios for which this approach gives 
a decision recommendation that is different from the optimal choice given by the far-sighted approach \citep{Alyaev2018b}. 
At the same time, the naive approach is superior to myopic optimization that only considers one step ahead which was used 
in previous papers with ensemble-based workflows, e.g. \cite{Luo2015,Chen2015spe}.
The DSS optimizes the complete well path ahead of the bit against the currently available 
representation of the geological uncertainty (as represented by the set of realizations).
Hence, the realizations should capture the complete current view on the geology with its interpretation uncertainties.
The next subsection summarises the implementation of the algorithm.

\subsection{Real-time ensemble-based optimization algorithm}
For simplicity 
we assume near-horizontal drilling and therefore we can associate the decision points $D_k$ with their position $x_k$ along 
the horizontal axis.
At every $x_k$ we discretize the trajectory alternatives by the well depth and denote the depths as
$z_{i}$ for the horizontal location $x_k$. 
Moreover, to account for the dogleg severity constraint (see the Appendix), it is important to take the current well angle $\alpha_{i_k}$ into account.

\begin{figure}
 \centering
 \includegraphics[width=0.5\textwidth,trim={0cm 10cm 20cm 0cm},clip]{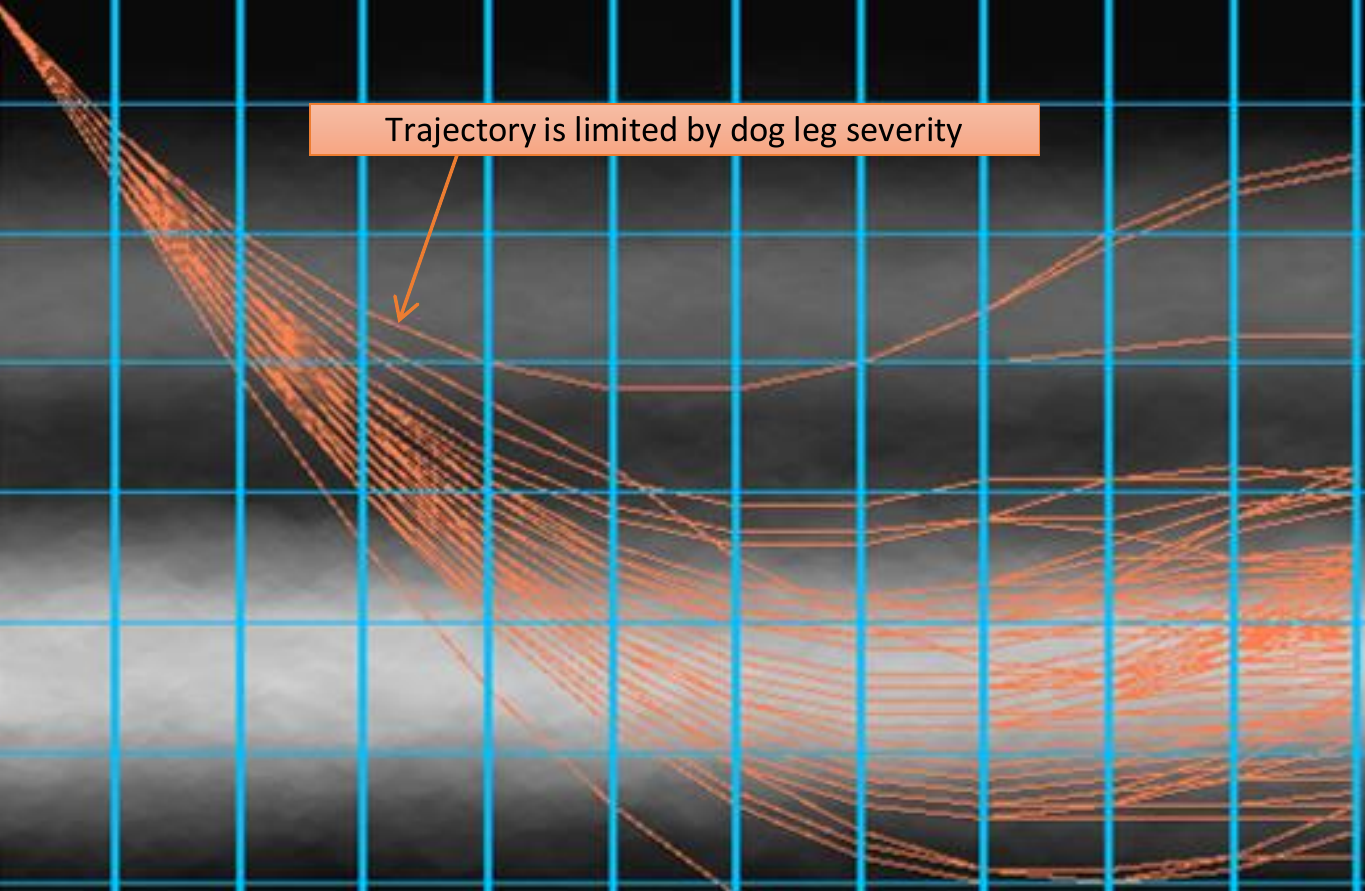}
  \caption{An example of discretization of trajectories. Vertical lines correspond to $x_k$ 
while horizontal lines correspond to $z_{i}$ (every 10th line in the set of possible depths is displayed). The orange 
polylines are possible well trajectories that go through the decision-grid points. One can see that the 
well trajectories are constrained by the dog leg severity; unreachable 
trajectories are not considered.}
 \label{figTrajectories}
\end{figure}

The decision for step $D_k$ at $(x_k, z_{i_k}, \alpha_{i_k})$ is either to stop or to add a segment connecting to a point $(x_{k+1},z_{i_k,next})$.
The choice of depth $z_{i_k,next}$ is constrained by the dogleg severity given by the user input, which in a discrete sense is approximated by 
$\alpha_{i_k,next}(x_k, z_{i_k}, x_{k+1},z_{i_k,next}) - \alpha_{i_k}$. 
The optimization algorithm evaluates  different trajectories that are represented 
as piecewise linear curves that go through the points $(x_k,z_{i_k})$ ahead of the current decision point, see figure 
\ref{figTrajectories}.
The resolution of points can be decided by 
the user and will affect the trade-off between the optimality of decisions and the computational time.

\todo[inline]{fixed? rev1:
Page 14 Line 308: Can you explain more about the following phrase "corresponding to the stopping decision for cases where the objective function after a certain point is always negative"? When will the algorithm choose to stop drilling, when the expected value of objective functions for all possible alternatives are negative or when the objective functions for all realizations are negative?}

For the decision at any decision point $D_k$, the decision algorithm consists of 
two steps.

In the first step, a dynamic programming 
algorithm finds a deterministic optimal well path for each realization and for every starting point by iterating over all possible trajectories.
The result of this step is the set of optimal decisions for every point-angle pair $(x_k, z_{i}, \alpha_{i})$ for every realization $j$ expressed as a $z$-coordinate for $x_{k+1}$ (or "stop drilling"
\footnote{The "stop drilling" decision means that for the current realization a positive value cannot be achieved, and stopping is the best alternative.
}):
\todo[inline]{ECS: I think that labelling a single equation with two equation numbers looks weird. But I see that both labels are used in the text. So maybe it is possible to define each of the two expressions in separate equations (assign a variable to each (A, B)), then define hat zi in terms of A and B? 
\\
SEAL: I agree, weird, but serves the purpose. I'll keep as is and see what correctors say.}
\begin{align}
  \hat z_{i,next}^j(x_k,z_i, \alpha_{i}) = 
  \arg\max_{z_{l} \textrm{ within constraints}} 
    &O([(x_k,z_i),(x_{k+1},z_l)]|M_j) 
    \label{eqSingleOptimal}\\
  + &\gamma O(X_{(x_{k+1},z_l,\alpha_l)}|M_j), 
  \label{eqSingleOptimalRecursive}
\end{align}
where $[(x_k,z_i),(x_{k+1},z_l)]$ denotes the next segment of the trajectory, $gamma$ is a discount factor, and $O(X_{(x_{k+1},z_l,\alpha_l)}|M_j)$ is the highest possible objective value that can be achieved for model $M_j$ from the trajectories $X_{(x_{k+1},z_l,\alpha_l)}$ starting with point $(x_{k+1},z_l)$ and angle $\alpha_l(x_k,z_i,x_{k+1},z_l)$.
\todo{ECS: confusing: is zi a discrete set of possible depths as indicated in Figure 4 above, or can zi be any depth? \\
SEAL: zi is a discrete set. Anything else written?}
In \eqref{eqSingleOptimal} the index $k$ is used to refer to horizontal locations, $i$ and $l$ for vertical locations and $j$ for the realizations respectively, as before.

\todo[inline]{ECS: if the discount factor 'gamma' is important, it should be explained here. Only referring to another paper makes this paper difficult to read. If it is not important, it could be deleted.}
\todo[inline]{SEAL: Erich, check out.}
The discount factor $0 < \gamma \leq 1$ in (\ref{eqSingleOptimal}-\ref{eqSingleOptimalRecursive}) is commonly used in the formulations of sequential decision problems \citep{feinberg2012handbook}.
It reduces (discounts) the value of decisions that are further ahead $O(X_{(x_{k+1},z_l,\alpha_l)}|M_j)$ compare to the immediate expected reward $O([(x_k,z_i),(x_{k+1},z_l)]|M_j)$ from the current decision.
In most of the paper, if not stated otherwise, we will use $\gamma = 1$ corresponding to the naive optimistic policy described in Section~\ref{secSecventialDecision}.
Values of $\gamma$ slightly less than one allow to reduce the value gained from the trajectory far ahead and thus present a practical way to compensate for the assumption of perfect information in the naive optimistic policy.

Equation \eqref{eqSingleOptimal} needs to be solved for the current position of the drill bit.
All the paths ahead are then recovered by finding the term \eqref{eqSingleOptimalRecursive} recursively.
In our implementation we follow the principles of dynamic 
programming  \citep{Cormen2009} to ensure that each point is evaluated only once and then tabulated to be reused in the other trajectories.
Thus the reconstruction of the optimal trajectories as well as the subsequent  evaluations for the objective function for the well is almost instantaneous.
In this way the optimal trajectory for 
each realization can be recovered:
\begin{equation}
	\hat X^j = [(x_0,z_{i_0}),(x_1,z_{0,next}^j),(x_2,z_{1,next}^j),...],
	\label{eqOptimalTrajectorySingle}
\end{equation}
where $(x_0,z_{i_0})$ is the starting point for the current optimization.
Similarly, by substituting \eqref{eqOptimalTrajectorySingle} into the objective function \eqref{eqGlobalObjective}, one can calculate the predicted well value for a given geological scenario.

In the second step, we need to perform a robust optimization to arrive at the single optimal 
decision: i.e. chose to "stop drilling" or steer towards the depth $\hat z_{0,next}$, whichever gives the 
best outcome on average, considering all realizations.

The computation of  $\hat z_{0,next}$ 
considers immediate permissible alternatives, including "stop drilling" and all $(x_1,z_{l})$ which are within the constraints of the dogleg severity, and
choose the one that is the best on average:
\TODO{similarly to eqref eqOptimization robust optimization equation}
\begin{align}
   \hat z_{0,next} = \arg\max_{z_{l} \textrm{ within constraints}} \sum_{j=1}^{n} \psi_j 
      &\left\{
   O([(x_0,z_0),(x_{1},z_l)]|M_j) 
   \right.
   \label{eqOptimizationGlobal}\\
   +&\left.
    \gamma O(X_{(x_{1},z_l,\alpha_l)}|M_j)
   \right\},
      \nonumber
\end{align}
where $\psi_j$ is the probability of realization $j$ and the rest of the notation is the same as in \eqref{eqSingleOptimal}. \footnote{Note that the evaluations for all the trajectories for the individual realizations 
have already been performed and cached on step one of the algorithm by applying equations \eqref{eqSingleOptimal}.
Also note that due to differences between equations \eqref{eqSingleOptimal} and \eqref{eqOptimizationGlobal} the final decision (steering or stopping) does
not necessarily coincide with any (depending on the realization) of the optimal decisions from step one.
}
We emphasize that equation \eqref{eqOptimizationGlobal} is used for exactly one decision ahead.
Thus the computational complexity of its evaluation is proportional to the number of immediate alternatives times the number of realizations. The vital consequence is that it does not suffer from the curse of dimensionality.
This distinguishes our optimization strategy from earlier approaches (e.g. \cite{Barros2015}),
which try to optimize all future decisions while neglecting the future learning.

The optimization algorithm presented in this section extends the classical robust optimization 
\citep{Chen2015spe,lorentzen2006new,chen2009efficient} 
to include the up-to-date knowledge 
when optimizing the full trajectory ahead of the bit. 
Due to the possibility of future learning, it is essential 
that only the first point is chosen by the robust optimization
while 
the rest of the trajectory is allowed to differ from realization to realization.
The future learning is expected to reduce the geological uncertainty and improve the decision for the next decision point.
The full well path joint optimization for the whole ensemble is costly and not always justifiable for a workflow where updates of the realizations are performed sequentially in time when new measurements arrive during drilling and time is scarce.
Instead, the decision for the next step (the next decision point) is recomputed 
once the new measurements become available and the ensemble is updated. 
This strategy allows for a real-time reaction to new information while also considering the 
prior information at every decision point.
From the perspective of  decision theory the strategy is equivalent to dynamic programming with assumption of perfect information.

\subsection{Visualization of the real-time modelling results}
\begin{figure}
 \centering
 \includegraphics[width=\textwidth]{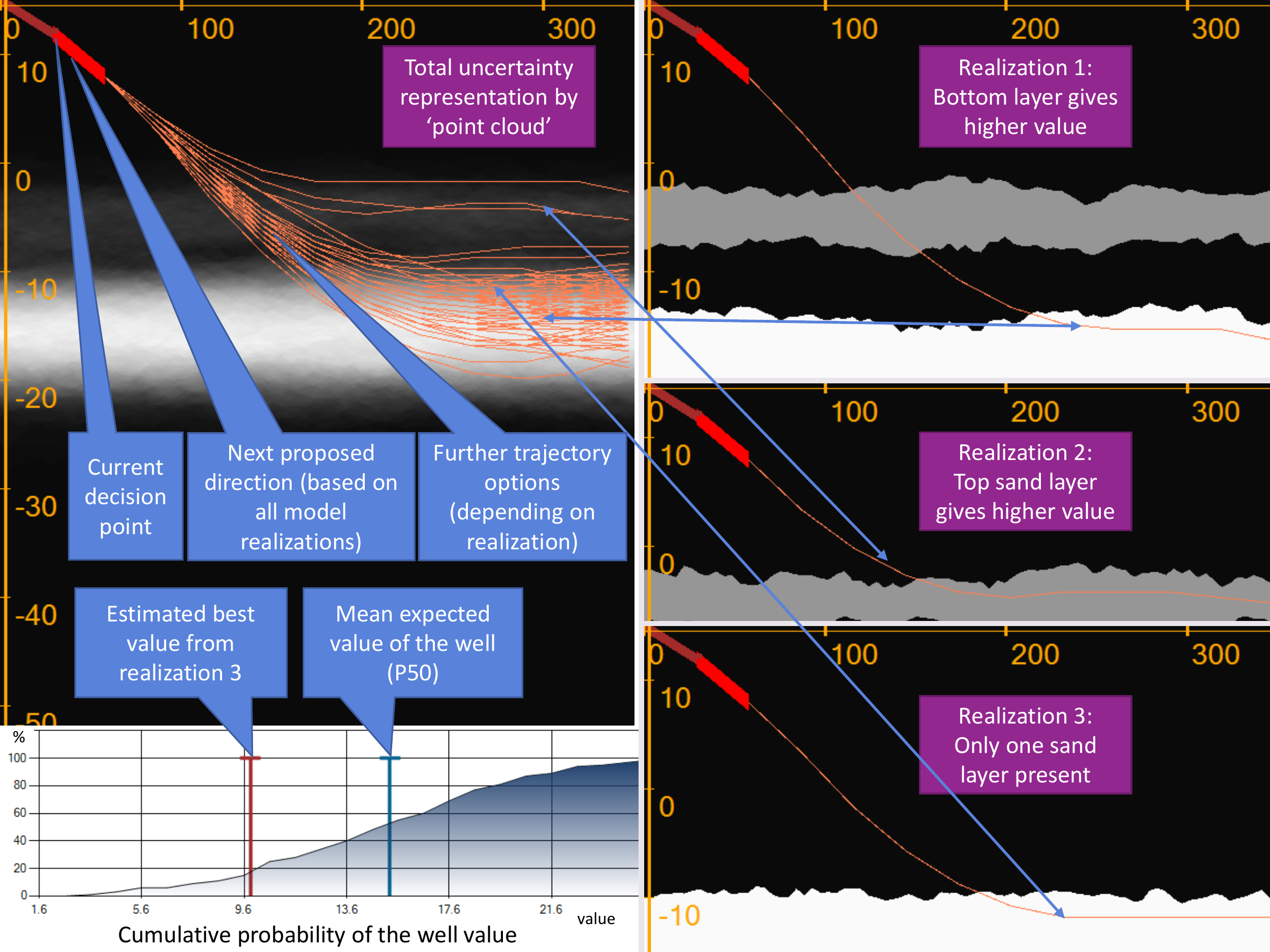}
 \caption{An illustration of the functionality of the DSS interface. 
To the right in the figure,
individual geomodel realizations are visualized. The realization to be visualized can be selected by the user.
To the top left in the figure, a 'point cloud' view of the total ensemble of realizations is shown.
The arrows indicate how the optimal trajectory for each realization is visualized in the 'point cloud' view.
To the bottom left, a cumulative diagram of the expected value of the well (including costs and future income) is presented.
It is based on the 
current uncertainty captured by the ensemble, with the vertical marks for (i) the mean expected value and (ii) the estimated value for a 
specific realization.
}
 \label{figVisualExample}
\end{figure}

The adoption of any DSS
requires that the system can be trusted by its users. 
Therefore the communication to the user of the reasoning behind the proposed decisions is 
essential. 
In the user interface, the proposed decision is visualized and the basis for the 
decision is explained. 
\TODO{ ERICH : This is somewhat misleading, in a bayesian framework each individual realization does not carry any more information than other realizations. And there is an infinite amount of possible realizations. If there is a value in examining a certain realization, it should be explained (maybe to get an overview, or a'better understanding of the results of the optimisation). If not, a reformulation is in order}

The main basis for a decision is the up-to-date probabilistic earth model.
In Figure \ref{figVisualExample}, an earth model with two oil-bearing sand layers with high 
resistivities
is used for the demonstration of the DSS visualization.
Between these two layers there
are background shales with low resistivity. High resistivity layers are indicated with a bright
color, while relatively low resistivity layers appear as gray. Black layers have very low 
resistivity and correspond to shale.
To the right in figure \ref{figVisualExample}, three (out of normally a hundred) realizations 
are shown.
In the user interface any realization can be selected for examination and the realization on
display can be effectively switched within milliseconds.
Moreover, the uncertainty can be visualized as a 'point cloud'. That is, for each point in space 
we visualize the average of the resistivity value over the ensemble of realizations as shown in in 
figure \ref{figVisualExample}.
In many cases, the point cloud is an intuitive way to understand the distribution of the uncertainty within the 
current ensemble. 

At all times
the interface highlights 
the consequence of the immediate decision (next proposed well segment)
in thick red and with a written communication of the calculated decision: an angle in degrees or 'stop'. 
The decision recommendation is supported by a cumulative plot of the expected value of the well based on the 
estimated geological
uncertainty, shown in the bottom left corner of figure \ref{figVisualExample}.
The plot should be interpreted as follows; for a selected value on the x-axis, the plot surface 
corresponds to a percentage, e.g. 20\%. That means that in 20\% of the realizations this value 
is not achieved. However, the value is exceeded in the remaining 80\% of realizations.

Furthermore, the interface communicates the two-step process behind the decision optimization as explained in 
the previous subsection.
When an individual realization is shown, the corresponding optimal trajectory resulting from 
optimization 
step one \eqref{eqOptimalTrajectorySingle} is visualized (starting from the uniquely selected next segment).
At the same time the value expected from this trajectory is marked on the value plot.
For convenience, the mean predicted value is also shown.\footnote{It is important to note that the 
mean does not necessarily coincide with a value expected from any realization.}
In the 'point cloud'  view, all the optimal trajectories corresponding to each
realization
are visualized (see figure \ref{figVisualExample}). 
By evaluating the density of trajectories, this latter display gives an intuitive understanding of the alternatives that are in reach of the 
current 
operation.

The display is fully updated with the new optimization results when the realizations are updated,
or if the user adjusts the objectives.
Further discussion of the flexibility of the graphical (GUI) and programming  (API) interfaces of the system  can be found in \cite{Alyaev2018}. 
In the rest of the paper we will use the point cloud view to visualize uncertainty and the reachable optimal trajectories for each realization to indicate the outcomes predicted by the system.

\section{Numerical examples}
In this section we demonstrate the performance of the DSS on synthetic examples.
In the examples we use a layer-cake earth model with two oil-bearing sand layers surrounded by 
background shales (see e.g. Figure \ref{figVisualExample}). 
\todo{ECS: No pinch-outs. Remove from conclusions.}
This seemingly simple model presents a challenging setting for making decisions:
the layer depths and thicknesses are uncertain and will be updated while drilling, and the drilling target is not predefined but is selected dynamically based on a multi-objective value 
function. 
The value function accounts for the estimated production potential of the well versus the 
estimated cost of drilling. 
Fast and consistent evaluation of this function while considering different alternatives 
under geological uncertainty
is the key to good decisions.

The operation starts 15 meters above the expected reservoir top 
(the expected top location is taken as zero) 
with an 80 degree inclination as commonly used directly before landing 
\citep{cayeux2018analysis}.
The operation is assumed  to end after drilling 350 meters in true horizontal length.
The decision points are equally spaced in the horizontal direction with approximately one stand between them  (28.6 meters apart).
The earth model is updated using unprocessed synthetic EM measurements (taken at the decision points) that are
modelled by
a simple integral model as in \cite{Chen2015spe}.
The depth of investigation (DOI) of the synthetic tool is about 5 meters, 
see figure \ref{figDepthInvestigation}. The data variance in the update equation \eqref{eqEnkfUpdate} is set to be 0.5 in dimensionless resistivity units. 

\todo[inline]{
Reviewer 3: The case of layered model is too simple. It seems to me that the authors tried to make a simple problem more complicated. I am not sure if the proposed model will outperform other geosteering models. There are some format problems as well.}

All the tests in this section follow similar assumptions about the layered model.
The layers in the model can be distinguished by their resistivity that is assumed to be known for
the synthetic cases. The resistivity values are set to 10 for shale and to 150 and 250 for the top and bottom sand layers respectively (all in dimensionless units).
The initial ensemble of realizations is created based on the expected boundary depths that vary around a mean value of 0, -5.3, -13.3 and -20.1 meters respectively. 
Depth uncertainties are generated using 
an exponential variogram model (nugget=0, sill=2.5, range=350m) following implementation from \cite{cressie1992statistics}.
Furthermore, co-kriging is used  to 
correlate the boundaries of the neighbouring layers (with correlation parameter set to 0.7), similar to \cite{lorentzen2017history}.
For rigorous testing, we let the synthetic truth also be a 
layer-cake model.
\todo[inline]{SEAL: removed the details about paramtrization. It is not crucial.}

For the DSS we will consider the following main objective;
to maximize the well exposure to the reservoir sands.
Instead of fixing the target the system should choose the sand layer based on its thickness, under the assumption that a thicker sand results in higher oil content and therefore better production. The two sand layers are otherwise considered to be equally good for drilling and production.
Additionally, the well should be placed in the upper part of a layer for improved production, and the drilling cost should be taken into account.
The trajectory is constrained by dogleg severity of 2 degrees and a maximum inclination of 90 degrees to allow for efficient gravel packing.
The precise mathematical definition of the individual objectives as well as their weighting is given by equation \eqref{eqPrimaryObjective} in the Appendix (Section~\ref{secAppendix}). For simplicity the value functions are scaled to "equivalent meters in sand drilled". 
That is, one \textbf{unit} corresponds to the net present value that can be produced from one reference well stand  positioned along a sand layer with reference properties of one-meter thickness.

\subsection{Optimal landing in different geological scenarios}
First we want to test the DSS workflow for different geological  scenarios.
All decisions automatically follow the recommendations from the DSS.

\todo[inline]{FIXED
Page 20 Table 1: The manuscript will be more complete if you can explain more about the objective function and how it was constructed. Equations in the appendix should do the job, in case any readers would like to replicate the simulation or make some changes in the objective function. Also, it's not quite clear how the value in the 'sweet spot' is doubled. In which scale are the values calculated?}

\newcommand{\halfwidth}{0.33\textwidth}

\begin{figure}
\centering
  \includegraphics[width=\halfwidth]{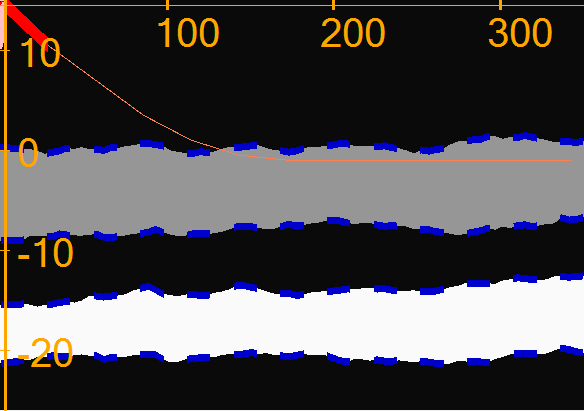}
  \includegraphics[width=\halfwidth]{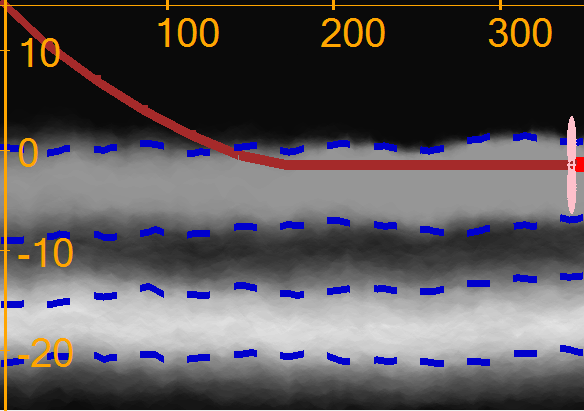}
  \includegraphics[width=\halfwidth]{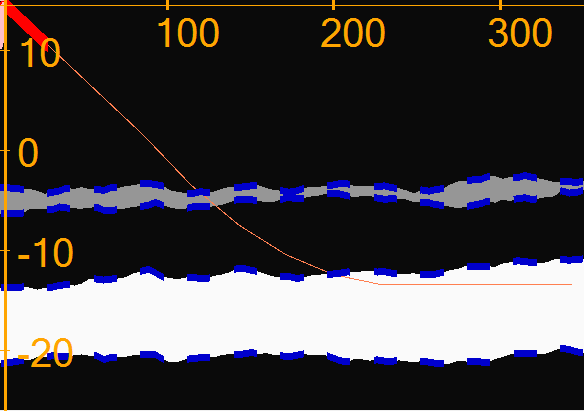}
  \includegraphics[width=\halfwidth]{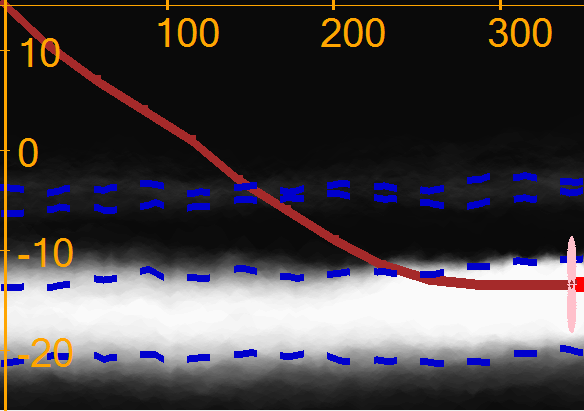}
  \caption{The synthetic truths for the two scenarios with their corresponding optimal trajectories (the two images to the left). 
  The value corresponding to each optimal trajectory is taken as the maximum theoretically possible for that scenario (100\%).
  The two images to the right show the final trajectories resulting from the application of the workflow in the two scenarios.
  Top scenario: The well almost matches the perfect trajectory and achieves 86.6\% of the theoretically possible value.
  Bottom scenario: The landing in the second layer is not perfect due to the initial 
uncertainty (not shown).
  Nevertheless the global optimization under uncertainty allows to adequately land the well in the 
bottom layer and achieve 58.6\% of the theoretically possible value.}
\label{figSteeringResult}
\end{figure}

To the left in figure \ref{figSteeringResult}, two different alternatives for the synthetic 
truth are considered followed by the results of application of the workflow on the figure's right. 
In the example we compare how the same set-up of the workflow operates step-by-step 
for these two different scenarios (everything is the same except for the synthetic truths).
The truth in both scenarios contains two reservoir layers. In the top scenario the top layer is thicker and 
hence more profitable while in the bottom scenario the bottom layer should be prioritized.
The images with the synthetic truths also contain the trajectories that are optimized with respect to the chosen metric (see equation \eqref{eqPrimaryObjective} in the Appendix).

We start from an initial ensemble of  realizations representing the model with uncertainty (the ensemble is the same for both scenarios). 
The 'point cloud' representation of the initial ensemble is depicted in the first column  in figure 
\ref{figStart}.
The blurry contours of the boundaries indicate the uncertainty in 
the layer positions and thicknesses. 
Initially, the global optimization 
foresees different decision outcomes 
that would result in 
landing in either the top or 
the bottom layer (the same for both scenarios as the pre-drill ensemble is the same, no model updates have been performed yet).
Therefore the DSS proposes the same initial decision for both scenarios: a build-up from 80 to 81.1 degrees, allowing for  future well landing in either of the layers.

In the next two decision steps (columns 2 and 3 in Figure \ref{figStart}) the tool look-around is insufficient to reach the sand layers (the sensitivity of the tool shown in pink).
Therefore no update takes place and, since the geomodel is the same,
 the steering decisions are the same for both scenarios. 
The DSS proposes an angle build-up 
that allows for better landing in the top layer, which seems more promising under the current view on the geological 
uncertainty. At the same time the alternative to drill to the bottom layer is not disregarded, as 
indicated by the optimal well paths in some of the realizations.

In column 4 
in figure \ref{figStart} the expected top boundary of the  top sand comes within the DOI of the tool. The uncertainty captured in the geomodel is correspondingly reduced after the update, rendering the top boundary sharper on the averaged image. 
At this stage  it is still uncertain  which layer 
that will be chosen for both synthetic truths,
but the objectives dictate to steer downwards in the bottom scenario to be able to reach
the bottom layer faster, if required.

In column 1 in figure \ref{figMid} the uncertainty in the top layer depth and thickness is reduced 
even further (notice the sharp boundaries of the top layer in the top image in column 1). 
More precise knowledge of the reservoir layers and the profitability of drilling in each of them puts more priority 
to landing in the top layer for the top row scenario.
Contrary to that, in the bottom row scenario the decision is to cross the shale between the two sands and drill to the bottom sand layer.
After the update performed in column 2 of figure \ref{figMid}, the uncertainties for the top layer boundaries are further decreased and the landing strategies are confirmed for the two scenarios.
These decisions are consistent with our 
knowledge about the truth in both scenarios. 

In columns 3 and 4 in figure \ref{figMid},
in the top scenario, the well is landed in the top layer. 
At the same time the DSS estimates that it might 
be better to drill downwards 
for some realizations. This is reflected in the figure by the thin well trajectories.
In column 4 in the bottom scenario in figure \ref{figMid} the DOI of the tool reaches the expected top boundary of the bottom layer. The uncertainty about the depth of the roof of the layer is reduced, yielding a more detailed landing plan.

The rest of the synthetic operation is shown in figure \ref{figFinal}.
In the bottom row of the figure, one can observe how the well is landed successfully in the bottom layer.

During the decision steps described in this section, a complex workflow consisting of the update 
loop and the DSS is running behind the scene.
The measured data is generated using the described EM acquisition model from the synthetic truth, including added measurement noise.
Every new measurement triggers an iteration of the update loop.
On a workstation with 20 logical cores a full model update takes 
less than a second for an early software implementation that is not optimized for production. 
Afterwards the DSS optimizes the trajectory of the well across all realizations and gives a result 
within another 10 seconds.

Figure \ref{figSteeringResult} shows the final step of the operation for the considered scenarios.
In both scenarios the DSS manages to land the well in the layer which is optimal with respect to 
the objective. We emphasize that this is possible due to the fact that
the full well trajectory is optimized against the up-to-date uncertainties.

In the scenario in the top row 
(figure \ref{figSteeringResult}), the steering result is close to optimal. This can be seen by visual 
comparison of the actually drilled well path to the left and the optimal well path to the right. The actual well achieves a value of 95.56 equivalent meters of reservoir sands drilled (with respect to objective \eqref{eqPrimaryObjective}), which corresponds to 86.6\% of the theoretically possible. 
Since much of the well length in our paper is used for landing, we cannot use a standard reservoir contact metric, but instead compare the value the theoretical maximum.
The theoretical maximum is the value that the
well can achieve when the trajectory is optimized with respect to the known true model (without uncertainty).
Obtaining complete information of the subsurface is not possible, therefore the theoretical 100\% will never be accomplished in practice.

\todo[inline]{ECS: next sentence: 'less likely' implies that the pre-drill uncertainty was not well captured in the geomodel. That should be commented, including that the updates handle this by updating the uncertainty estimations while drilling.
I've changed the text, but are you sure that this uncertainty was not captured in the initial model? I cannot see why it shouldn't have been ...}
The bottom scenario was generated with different parameters for the layer boundaries yielding a thinner top layer.
Thus it is a less likely scenario with respect to the pre-drill geomodel.
However, as the real-time data, which indicates a thinner top layer than initially estimated,
becomes available, the geomodel uncertainty is updated, the well path is corrected and the optimal target is reached. 
The well trajectory resulting from the bottom scenario has a value of 49.81 equivalent meters of sand drilled, which is  approximately 58.6\% of the theoretically possible value for that scenario.
Note that new developments improving the pre-drill model or the look-around/look-ahead 
capability (such as in \cite{constable2016looking}) will improve the decision outcomes 
provided by DSS.

Finally, we note that for both scenarios the reservoir boundaries are automatically mapped along the wellpath. The updated uncertainty estimations can thus benefit the further reservoir development planning.

\newcommand{\narrowwidth}{0.24\textwidth}
\begin{figure}
\centering
    \includegraphics[width=\narrowwidth]{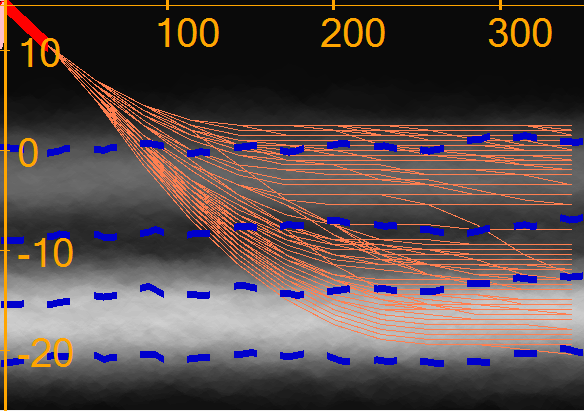}
  \includegraphics[width=\narrowwidth]{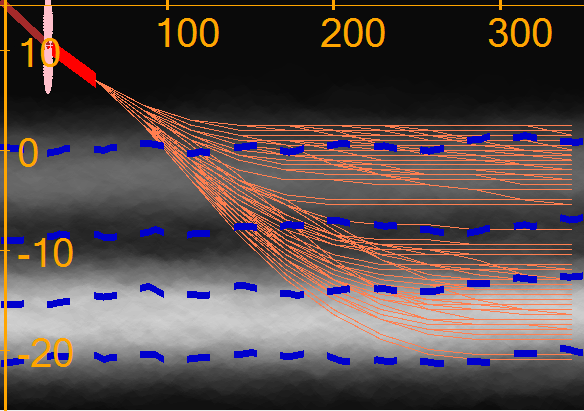}
  \includegraphics[width=\narrowwidth]{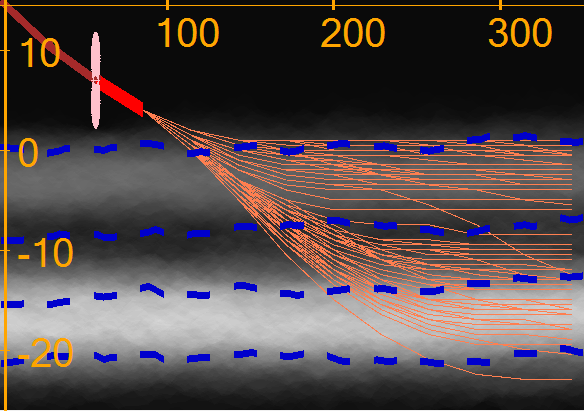}
  \includegraphics[width=\narrowwidth]{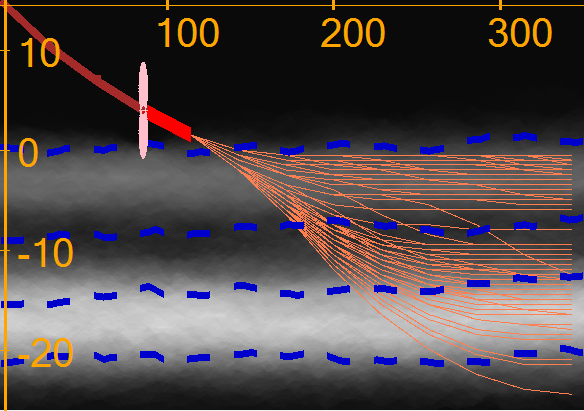}\\
      \centering
    \includegraphics[width=\narrowwidth]{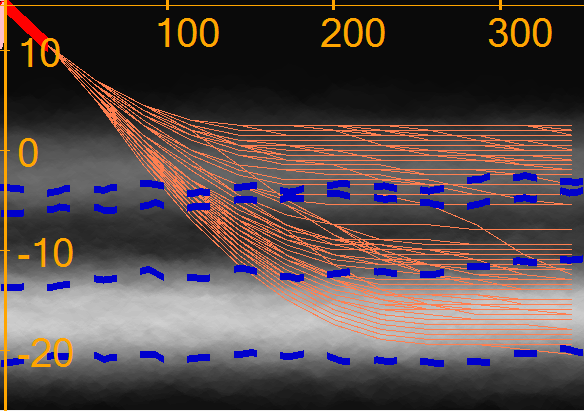}
  \includegraphics[width=\narrowwidth]{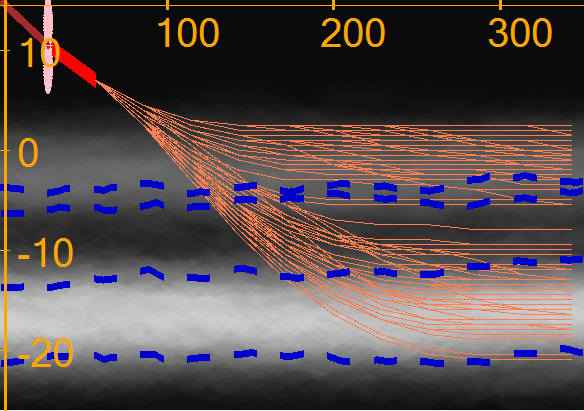}
  \includegraphics[width=\narrowwidth]{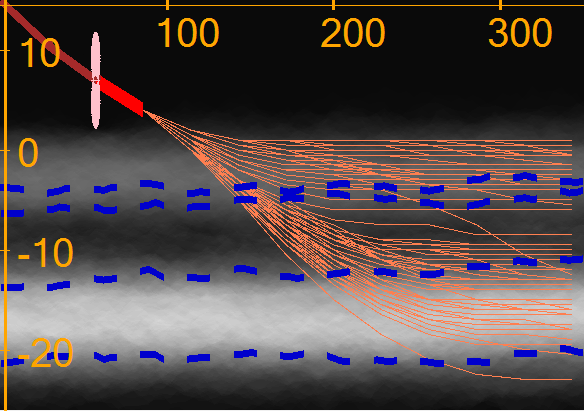}
  \includegraphics[width=\narrowwidth]{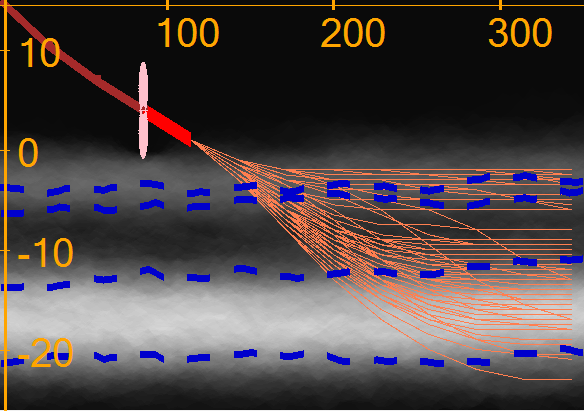}
      \caption{Demonstration of the decision support system  for the two synthetic scenarios in the top and bottom row as in figure \ref{figSteeringResult}. Blue dashed lines indicate the positions of the layer boundaries in the synthetic truths.
The figures show the step-by-step decision recommendations from the DSS (as indicated by the advancing bit).
The initial geomodel uncertainty is the same for both scenarios. 
}
\label{figStart}
\end{figure}

\begin{figure}
\centering
    \includegraphics[width=\narrowwidth]{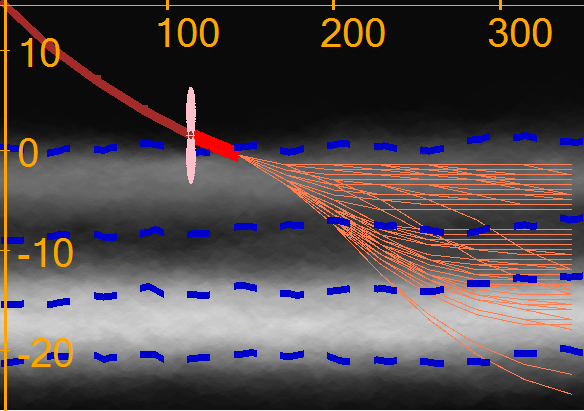}
  \includegraphics[width=\narrowwidth]{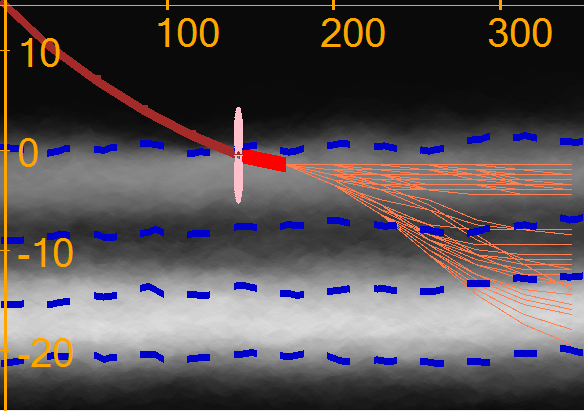}
  \includegraphics[width=\narrowwidth]{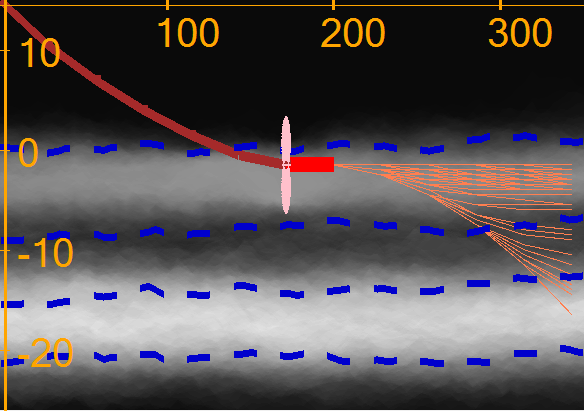}
  \includegraphics[width=\narrowwidth]{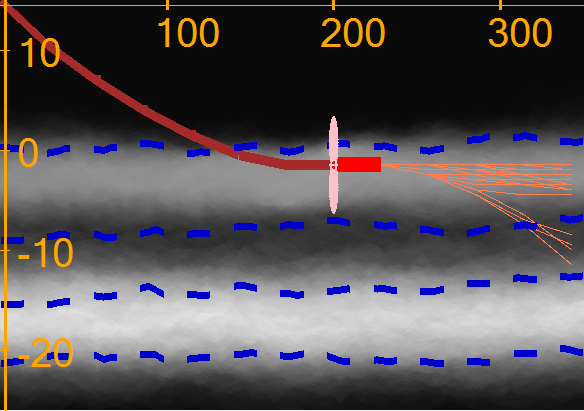}\\
  \centering
        \includegraphics[width=\narrowwidth]{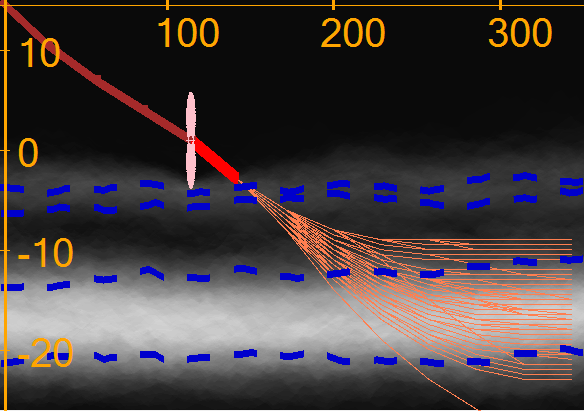}
  \includegraphics[width=\narrowwidth]{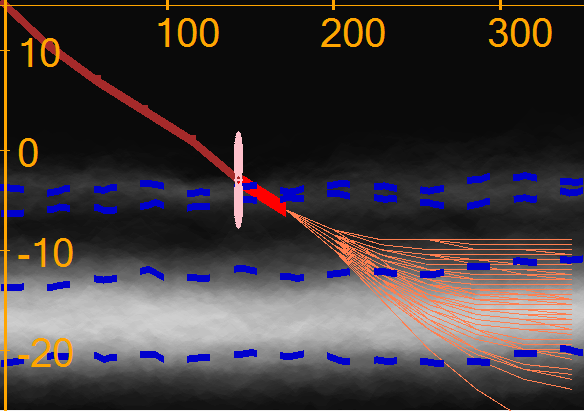}
  \includegraphics[width=\narrowwidth]{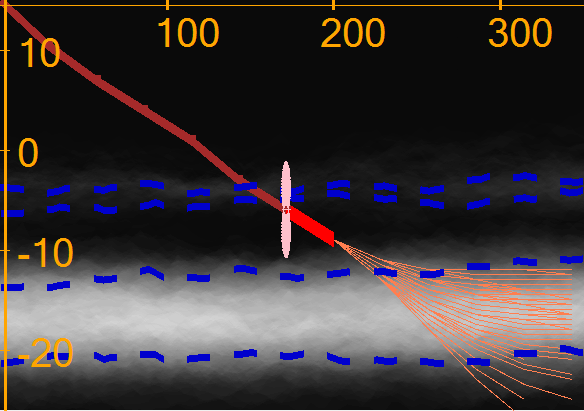}
  \includegraphics[width=\narrowwidth]{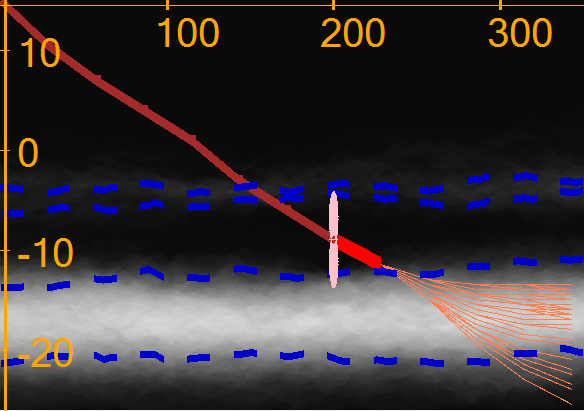}
  \caption{Continued demonstration of decision support system for two synthetic scenarios from 
figure \ref{figStart}. The figures shows the step-by-step outputs of the DSS.
As the drilling operation progresses and more data become available, the top layer is preferred for the top 
scenario and the bottom layer is preferred for the bottom scenario.
}
\label{figMid}
\end{figure}

\begin{figure}
\centering
      \includegraphics[width=\narrowwidth]{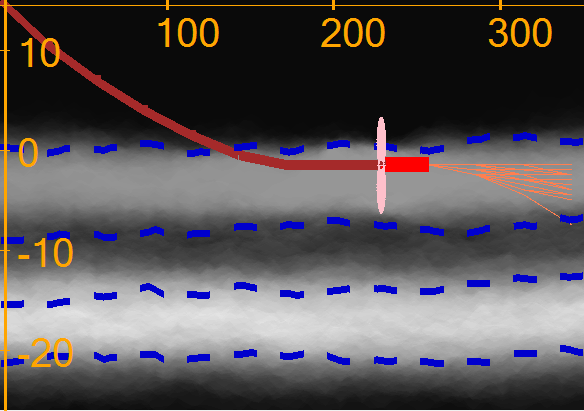}
  \includegraphics[width=\narrowwidth]{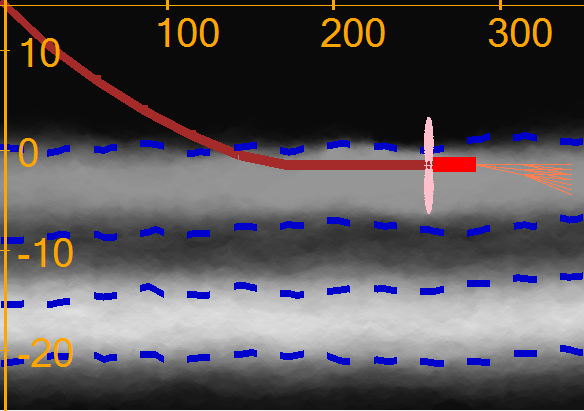}
  \includegraphics[width=\narrowwidth]{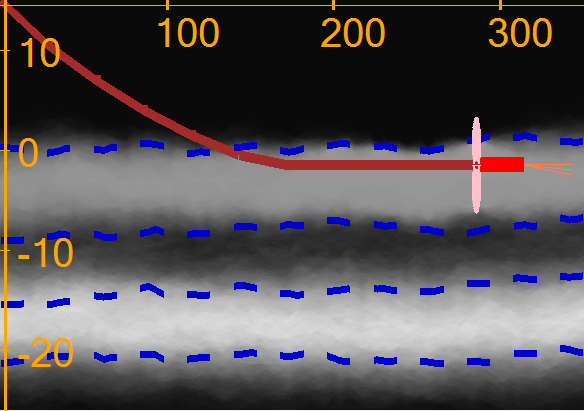}
  \includegraphics[width=\narrowwidth]{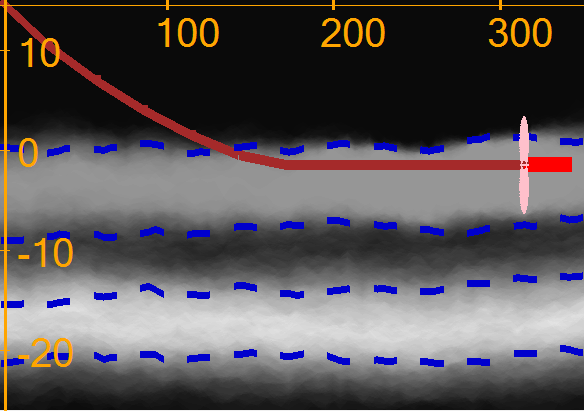}\\
    
        \includegraphics[width=\narrowwidth]{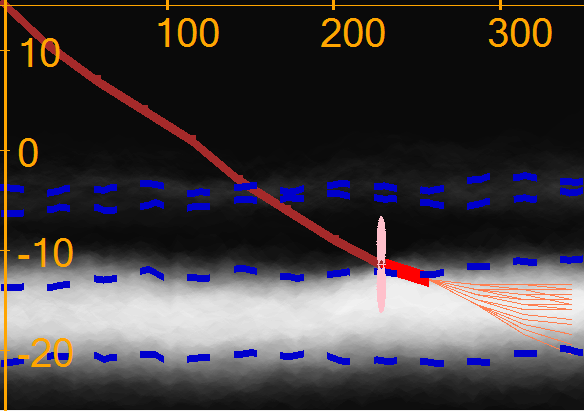}
  \includegraphics[width=\narrowwidth]{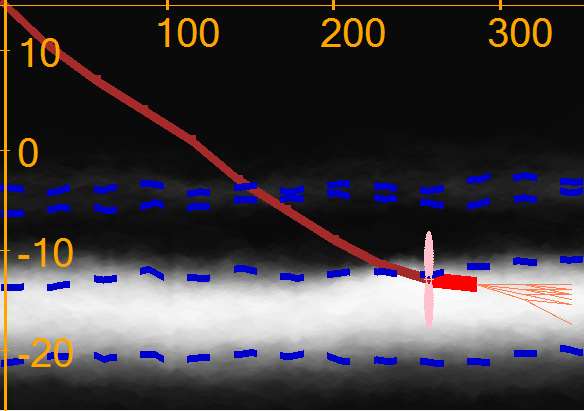}
  \includegraphics[width=\narrowwidth]{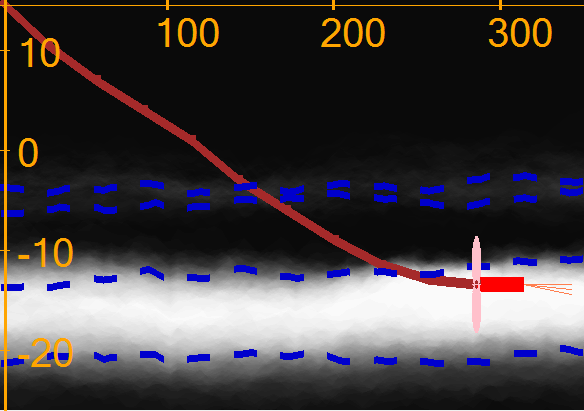}
  \includegraphics[width=\narrowwidth]{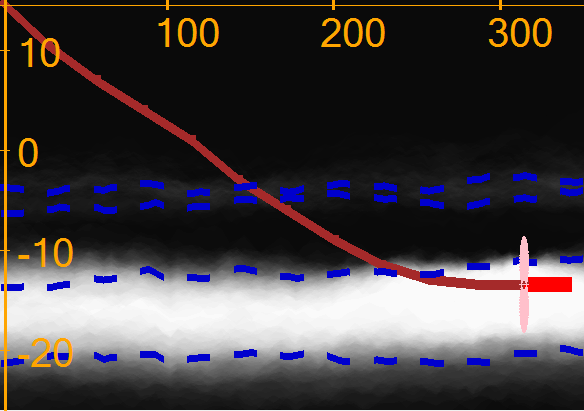}
      \caption{Continued demonstration of the decision support system for the two synthetic scenarios from 
figure \ref{figMid}. The figures show the step-by-step outputs of the DSS.
}
\label{figFinal}
\end{figure}

\subsection{Statistical analysis of the performance of the DSS}
\label{secStatisticalPerformance}

In the first numerical example we presented two situations where the DSS successfully landed the well in the optimal layer despite the initial uncertainty. 
Obviously, due to both the uncertainty of the subsurface interpretation and the simplifications in the DSS' "naive" algorithm, such good results would not be achieved for all cases.
In this example we investigate the statistical performance of the DSS.
To do so, we run 100 different synthetic geosteering cases. Their synthetic truths are drawn from the same distribution as used for the model realizations.
For all 100 cases, we follow the recommendations of the DSS with the same objectives as in the previous example (see \eqref{eqPrimaryObjective} in the Appendix).

To evaluate the DSS performance, we will look at two metrics;
\begin{enumerate}
    \item What is the value of the well resulting from the recommendations compared to the optimal well that is based on perfect information?
    \item Did the DSS land in the optimal layer? 
\end{enumerate}
\todo{TODO: verify my update in the list}
Both metrics are case-specific. 
To compare them to each other, we are evaluating them relative to the optimal well trajectory computed for the synthetic truth for each particular case (similar to figure \ref{figSteeringResult}). The value achieved by the well optimized with respect to the deterministic synthetic truth is set to 100\%.
The second metric is subjective as it is not directly included in the objective function, but it gives an intuitive operational understanding of the DSS' performance.
We let the 'successful landing' criterion be defined as drilling two complete drill-stands in the target layer (see bottom row in figure \ref{figFinal}).
As in the previous example, we finish the drilling operation after having drilled 350 meters in horizontal direction.

\todo[inline]{FIXED
Page 25 Line 521: Why two drill-stands is used as a success criterion? If the well lands further in the reservoir (away from 0 m on the x-axis), can we continue drilling the horizontal section until we reach the originally-planned well length?}

There are two challenges related to our objectives. For cases where the top layer is optimal, coming in at a low angle might result in overshooting the sweet spot which gives double the value.
For cases where the bottom layer should be preferred, the challenge is to realize early enough that the top layer is thin and drop the angle to get good coverage of the bottom layer.

\begin{figure}
    \centering
    \includegraphics[width=0.65\textwidth]{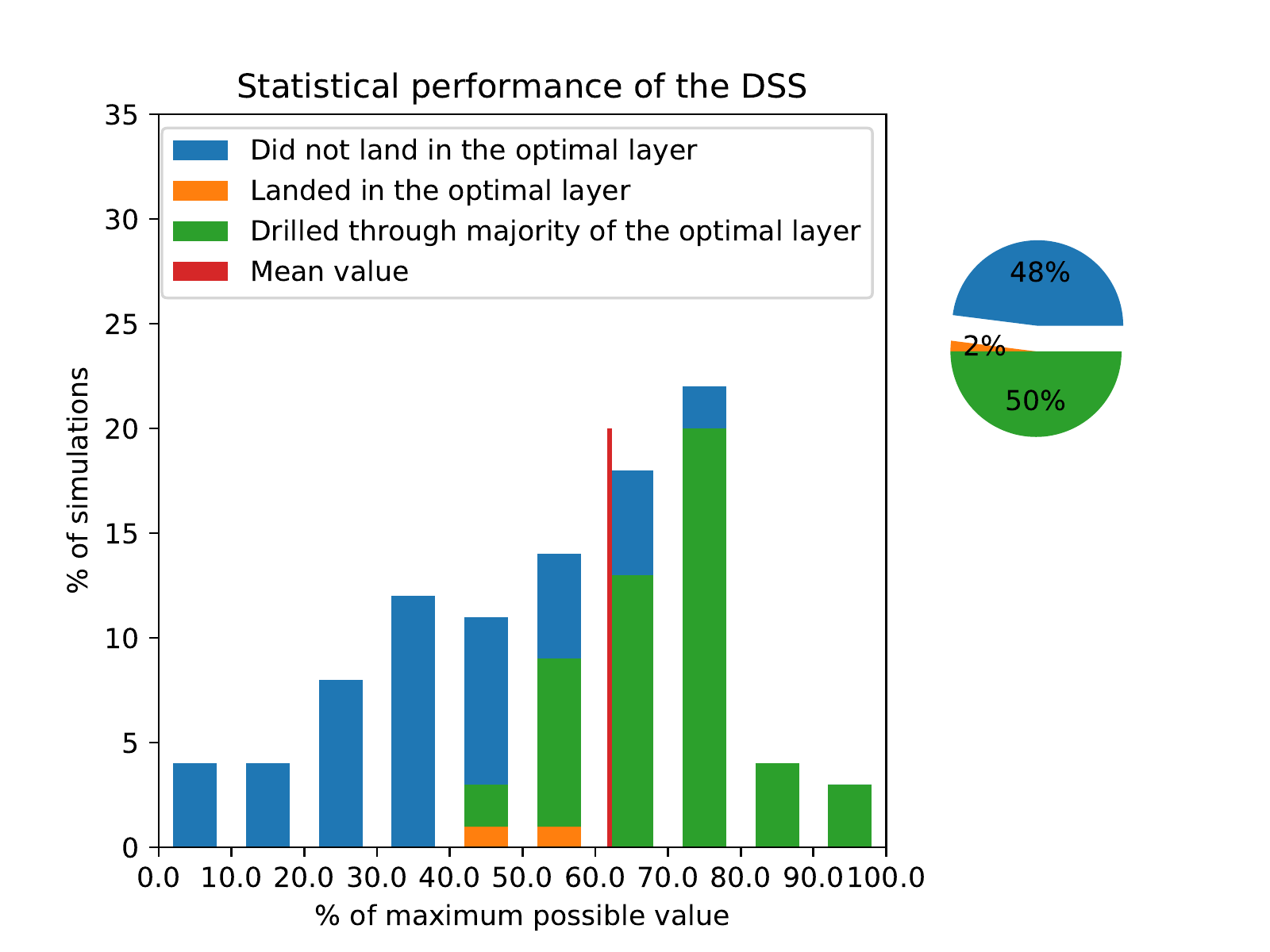}
    \caption{The results showing the statistical performance of the DSS. The cases are grouped in bins by the percentage of the theoretical maximum value achieved. The value of a resulting well is on average higher than 60\% of the case-specific theoretical maximum. 
        Good results are achieved even for scenarios where the well is landed in a sub-optimal layer.}
    \label{figStatResults}
\end{figure}

The statistics of the DSS performance is summarised on Figure \ref{figStatResults}.
The pie chart indicates that the well is landed in the optimal layer in 52\% of the cases.
Among those, in only 2\% of the cases the well length within the optimal layer is less than half of the maximum possible length.

We emphasize that the choice of layer was not explicitly included in the objective function. Thus, a fair performance evaluation is based on the value of the resulting wells. This is summarised on the bar plot in Figure \ref{figStatResults}.
The bars indicate the percentage bins of the maximum possible well value achieved by the DSS in each of the 100 different cases. 
The bars are split by the choice of layer for the resulting well. 
Not surprisingly, most of the better results correspond to the wells that landed in the optimal layer. 
At the same time, in the challenging geological conditions of the chosen setup, choosing the sand layer which is not optimal could result in over 70\% of the maximum possible well value (see blue squares in Figure \ref{figStatResults}).

\todo[inline]{FIXED?
Page 25 Line 533 (and other locations throughout section 4): What is the definition of 'maximum possible well value' or 'theoretically possible value'? In the manuscript, the value of the wells resulting from applying the method proposed here is measured as a percentage of maximum well value in horizontal well. Without a clear explanation, it could be easily confused with a 'reservoir contact' which is a value measure commonly reported in many geosteering papers with more inflated value around 90\% - 100\%. This could make the 60\% results in this paper seem less exciting. From what I understand, they are measured on different rules. If that's the case, you might want to state this clearly.}
\todo[inline]{ECS: I don't think that the reviewer's comment is well addressed yet (at least not in this section, haven't read the following sections yet). And I agree with him, if we don't use the same metric as is common in other papers, it should be explained. Looking only at the numbers show that the DSS is outperformed by manual geosteering (60 vs 90\%), and people might disregard out paper as not very important. Here are some proposals that may or may not be relevant;
1. Much of the well length in our paper is used for landing, so it will draw down the 'reservoir contact' percentage. If we drilled a 2 km well instead of 350 m, the 'reservoir contact' percentage should increase a lot.
1.5: I assume our metric is in fact different from 'reservoir contact', this should be explained (see the next items, and more if you come up with something)
2. 'Reservoir contact' in other papers means that the rest of the well has zero profitability. Is the same true for us, or is there also a value in drilling in shales?
3. Drilling cost is not included in traditional geosteering papers, but should influence our results.
4. 'Reservoir contact' in other papers is an interpretation (biased, may be inaccurate, etc)
ECS recommendation: I guess 1. is the most important, and should be commented as e.g. 'Our results are not directly comparable to the '90\% reservoir contact' often reported in other geosteering papers. This is because (see the explanation in 1.) '
}
\todo[inline]{SEAL: It is sad that it is not clear by now, but we need to stick to the chosen metric. I do not see a way to communicate it better than currently shown in the paper.}

The average value of the wells drilled by the DSS is 62\% of the theoretically possible value, see Figure \ref{figStatResults}. 
We reiterate that the 100\% value 
is impossible in a non-synthetic case since it requires knowledge of the true subsurface.

\subsection{Statistical performance of the DSS with a discount factor}
\label{secDiscountFactor}

\begin{figure}
    \centering
    \includegraphics[width=0.65\textwidth]{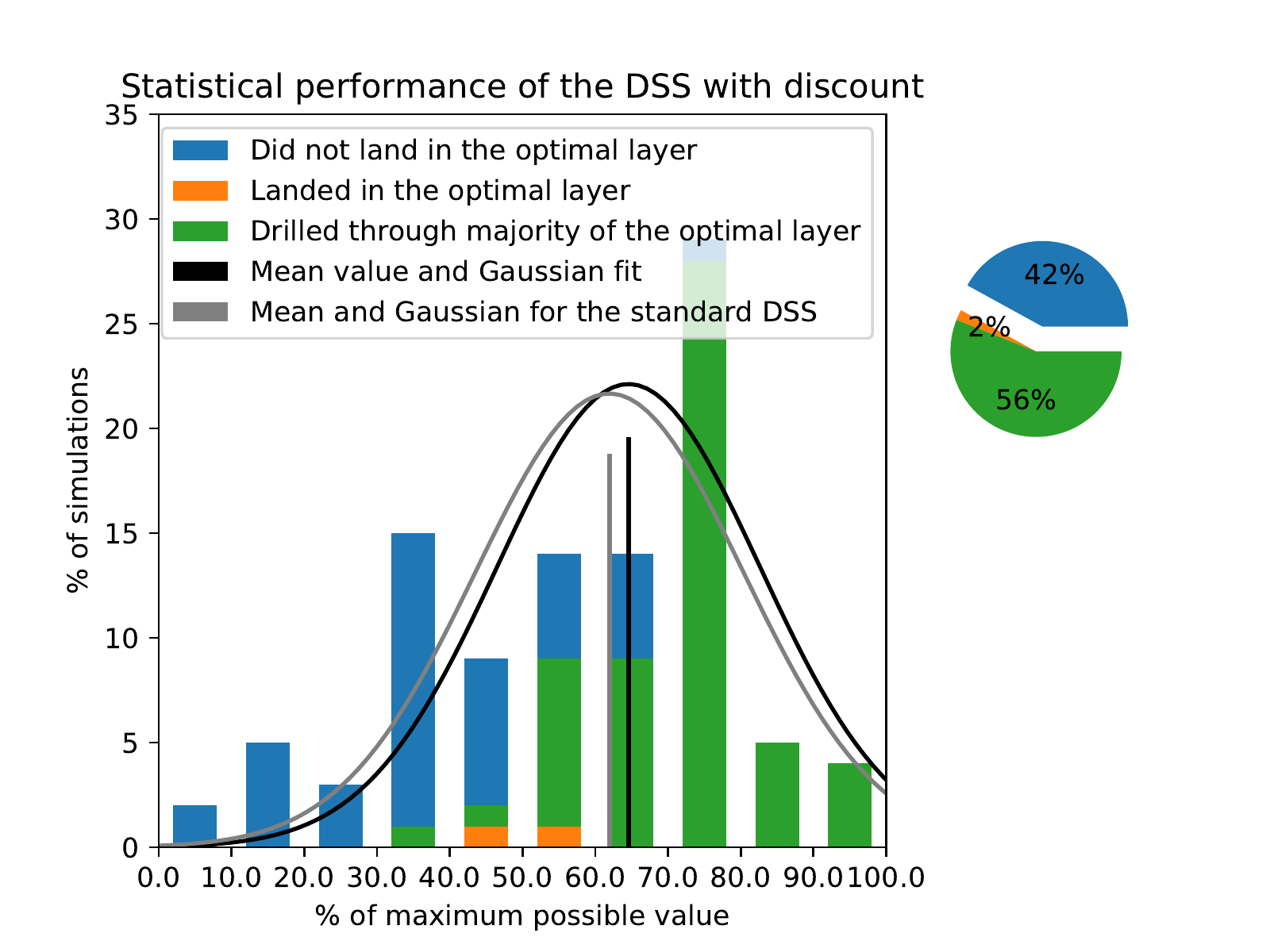}
    \caption{The results showing the statistical performance of the DSS with the discount factor of 0.9 to compensate for future learning. 
    The correction for future learning allows to achieve better results than in Figure \ref{figStatResults} (indicated in gray).}
    \label{figStatResultsDiscounted}
\end{figure}

Based on the theory described in section \ref{secSecventialDecision}, we know that the naive decision policy used for the DSS might not give optimal results as it neglects the modelling of the future learning \citep{Alyaev2018b}. 
One practical method to compensate for this is using a discount factor for future value 
$\gamma < 1$ in the optimization equations \eqref{eqSingleOptimal}, \eqref{eqOptimizationGlobal}.
The effect of a discount factor is hard to analyse theoretically, therefore we try to apply a naive decision policy with a discount in the following numerical example.

Here we set $\gamma = 0.9$ while keeping the rest of the parameters the same as in the previous example (Section \ref{secStatisticalPerformance}). 
The statistical results achieved by the modified DSS are presented in Figure~\ref{figStatResultsDiscounted}. Compared to the DSS without the discount factor, the average value achieved is increased from 62\% to 64.6\% with similar spread in the fitted Gaussian distribution. We also observe the increase of the "optimal" landings by 6\% compared to figure \ref{figStatResults}.

This example indicates that the DSS performance can be improved by introducing a discount factor for future well value. We expect that the optimal choice of the discount factor would depend on the selected application case of the DSS. Thus, the purpose of this example is to indicate this practical option in the DSS and not to find the optimal value of $\gamma$.

\subsection{Adjusting objectives due to insights}
\label{secInsights}
\todo[inline]{SEAL: continue here}
In the final example
we demonstrate the flexibility of the DSS by changing 
the weights of the different objectives.
This might be important during adoption in the field, as the insights gained during an operation might change the prioritization of the geosteering objectives. 
The main reason for changing the objectives and their weighting is the fact that for geosteering operations, the objectives are often simplified to ensure the possibility to evaluate them in real time. As more insight is gained, it may become clear either via expert judgment or various types of calculations or simulations that these simplifications and the initial weights may not yield the best decision suggestions (the specific considerations made to obtain better objectives and weighting is not in the scope of this paper).
The objectives and their weights can be adjusted in the user interface of the DSS.

\todo[inline]{SOLVED
Page 26 Section 4.3: How are the objectives adjusted during the operation? Is it resulting from simulation/calculation or expert judgments? For the purpose of this manuscript, I don't have any problems with either method, just to clarify which one is being used. }

\todo[inline]{ECS: it is not easy to come up with a reasonable example, then we need help from a subsurface expert. Whatever example I can think of, it is clear that it should be taken into consideration in the initial set of and weighting of objectives. This is why I phrase it like this. The real explanation for changing objectives is in line 459 above. Anyway, the new formulation is better than the earlier.}
\todo[inline]{ECS: another reason for changing objectives could be different phases in the operation. Landing may have its own objectives, just as drilling the horizontal section, or when drilling out of the reservoir to gain more information, or when drilling in the outskirts of the reservoir. Unless they should be mixed in a single complicated objective function...}
In this example we recall the operation described in the top scenario of the figures \ref{figStart}--\ref{figFinal}. 
Here we consider a slightly different drilling scenario, but using the same setup including the 
synthetic truth.
In column 2 in figure \ref{figMid}
the bit enters the top sand layer, which should 
result in a landing as shown in figures \ref{figMid} and \ref{figFinal}.
For the sake of argument, let us assume that real-time measurements indicate that the top sand is of poor quality. 
Thus, the geosteering experts take the decision to prioritize sand quality, which was not part of the original objective.
Consequently, the weight of staying in the top part of the reservoir is decreased to 0.3 and the weight of the 
sand quality objective is introduced and set to 0.7, resulting in a new global objective function described by \eqref{eqAlternativeObjective} (see the Appendix).

\begin{figure}
 \centering
 \includegraphics[width=0.98\textwidth,trim={0 2cm 8.2cm 0},clip]{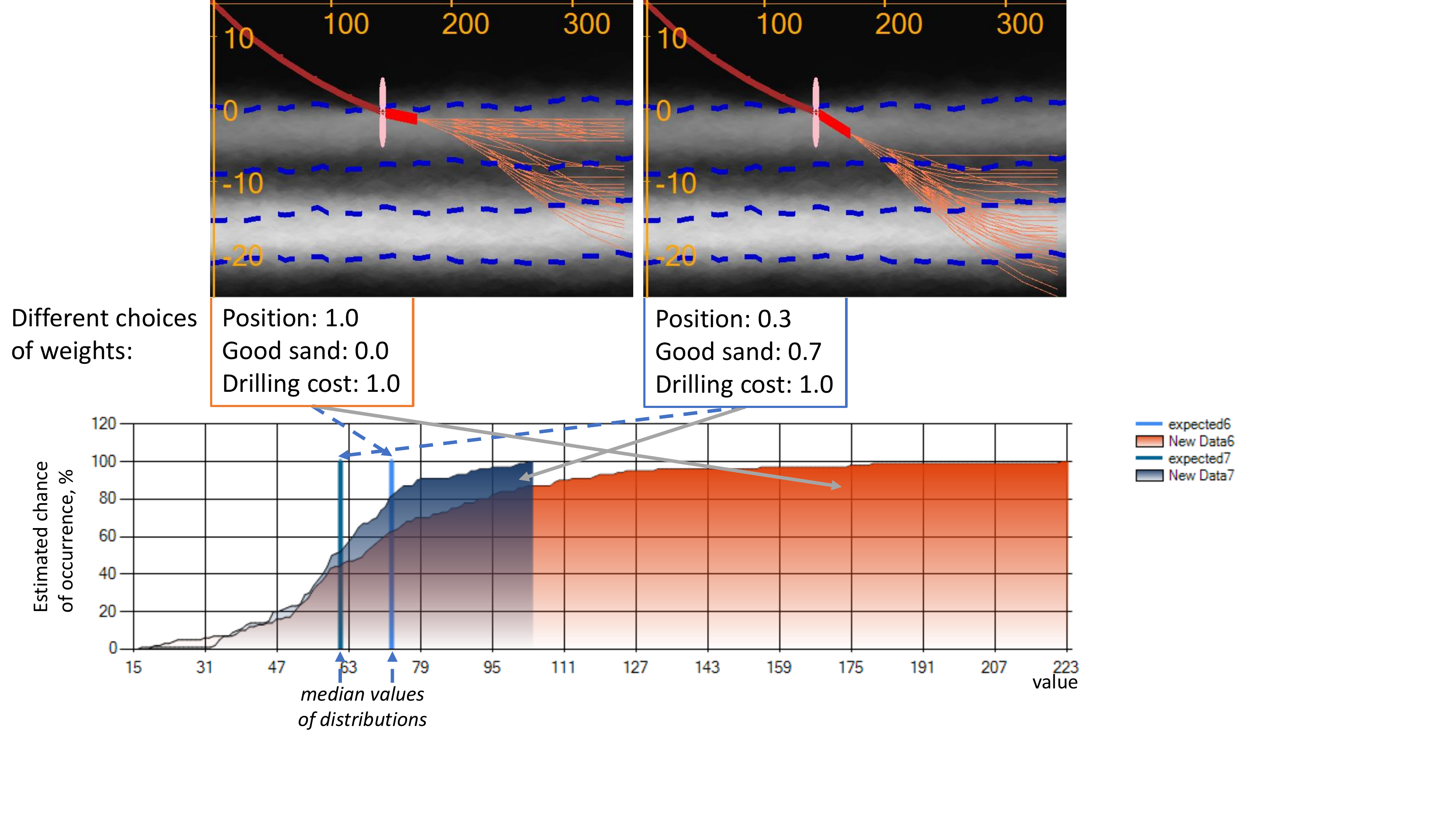}
 \caption{Example of how the weights of the objectives can be adjusted in the middle of the 
geosteering operation, in accordance with the insights gained while drilling (see the Appendix for a detailed description of the objectives).
The ensemble of realizations is the same in both
scenarios. 
The bottom plot compares the cumulative distribution of the resulting multi-objective value functions 
for 
the  choices of the weights (the vertical lines indicate the mean expected value for each distribution).}
 \label{figNewWeights}
\end{figure}

\begin{figure}
\centering
  \includegraphics[width=\halfwidth]{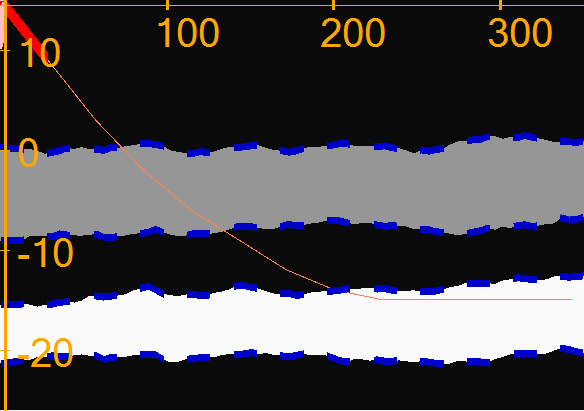}
  \includegraphics[width=\halfwidth]{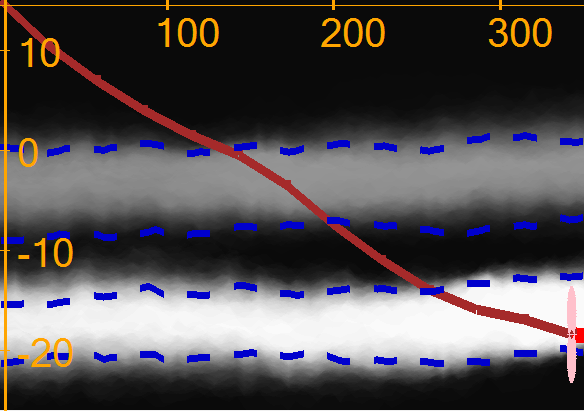}
  \caption{The final trajectory (right) for the scenario where the weights have been changed to pursue the 
layer with better quality sand following figure \ref{figNewWeights} (middle).
  For comparison, the optimal trajectory for the synthetic truth is shown to the left. 
The 
latter is computed applying the new  objectives \eqref{eqAlternativeObjective} from the start of the operation.}
\label{figAlternative}
\end{figure}

The newly selected weights can be applied to the trajectory optimization in real-time. 
The expected 
outcomes are shown in Figure \ref{figNewWeights}. 
The cumulative value diagram shown in the figure clearly indicates the reduction of 
the expected well value after the new objective function is chosen. 
The alteration in the objective 
 results in a landing in the lower reservoir layer, as shown 
in 
figure \ref{figAlternative}. 
Comparison of the optimal trajectories with their actual outcomes in figures \ref{figSteeringResult} 
(bottom) and 
\ref{figAlternative} gives a visual proof that superior results can be 
achieved when 
the 'correct' objective is selected before the start of an operation.

% challenge to compute the function. }
 
\section{Conclusions}
In this work we have presented a functioning consistent 
decision support system (DSS) aiming at 
supporting real-time geosteering decisions. 
The DSS provides directional drilling decision support 
information and recommendations. 
The recommendations account for real-time measurements behind and 
around the bit, inferred uncertainties ahead-of-the-bit, and multiple objectives. 
The system 
includes a visual display that allows the geosteering team to inspect uncertainties and immediately 
see and evaluate the possible results of their decisions. 
In contrast to basing geosteering decisions on "educated guesses" about the geological interpretation, future profit and drilling costs,
 the DSS provides a consistent Bayesian
framework for making ahead-of-the-bit inferences based on prior information and learning (from real-time data) 
while drilling.

The workflow implementation presented here can consistently update uncertainties ahead of the drill bit and provides a visual and interactive means to inspect the resulting multi-realization model of the subsurface.
However, this uncertainty quantification is not an end by
itself.
Rather, the goal is to make good 
geosteering decisions, which requires an assessment of relevant and material uncertainties. 
An essential part of improving the results from geosteering operations is to move the focus away from 
real-time data to actual decisions \citep{Kullawan2014}. 
The DSS uses real-time data gathering and 
learning-while-drilling to optimize key drilling decisions, thus ensuring good utilization of new measurement technologies.

There is abundant research and literature demonstrating that people are exceptionally bad at making 
decisions in complex and uncertain environments (see e.g. \cite{tversky1974judgment}).\footnote{The most recent Nobel Memorial Prize in 
Economic Sciences was given to Richard Thaler  for his contributions to behavioral economics. His 
former collaborator, Daniel 
Kahneman was awarded the same prize in 2002 for his work on the psychology of judgment and 
decision-making.}
The DSS embeds a consistent 
uncertainty quantification and a sophisticated decision-making process, and is particularly advantageous for unbiased 
high-quality decision support when navigation in complex reservoirs with several potential targets 
and significant interpretation uncertainty. 
The real-time performance of the system is of major importance for geosteering where time for 
evaluation, re-consideration and decision-making is scarce.

To illustrate the benefits of the DSS we have presented
synthetic cases with multiple objectives, for which the full workflow consisting of the model updating and the decision 
recommendations was applied. 
The system demonstrated landing in the reservoir,
automatic choice of target in a multi-target geological scenario,
and navigating the well in a layer-cake geological configuration.
results were consistently achieved  in several distinct scenarios as well as in a statistical test.
Statistically, the system-recommended decisions are initially achieving more than 60\% of the theoretically possible well value despite 
the uncertainty in the pre-drill and while-drilling geological interpretations. We expect that the performance will increase with future improvements of the system.
\todo{ECS: I added the last sentence to emphasize that these are initial results, and as a response to one of the reviewers comments}

Moreover, we have illustrated the flexibility of the implementation of the DSS.
The possibility to introduce correction for future learning gives a further average improvement of 2.6\% for the statistical performance of the system in our examples.
There is also a flexibility 
when it comes to 
adjusting decision objectives. 
By design the DSS reacts to changes in the objectives and constraints within seconds, providing  
unbiased decisions for the modified choices.

This paper presents proof-of-concept testing of the DSS.
The system uses existing measurement and modeling tools and identifies the optimal decisions 
through multi-objective optimization under uncertainty. It can be naturally extended to the advanced measurement technologies used in the field as well as include more realistic geology in its multi-realization geomodel.
With that, we see testing on historical operations as a possibility in the nearest future.

\section{Appendix: formal definition of objective functions and constraints}
\label{secAppendix}

The appendix describes the objectives and constraints used in the paper. First we define individual constraints and objective functions that have been used in the numerical examples. Thereafter, we define the weights that are selected to form the global objective function \eqref{eqGlobalObjective}. 

\subsection{Constraints}
In all examples we use the following two constraints:
    \begin{enumerate}
        \item The trajectory is constrained to a dogleg severity of maximum 2 degrees between decision points (approximately every 29 meters),
        \todo{SEAL: corrected to better match implementation}
        which is approximated as 
        \begin{equation}
            \left| \alpha_{i_{k+1}} - \alpha_{i_k} \right|
            \leq 2 \, \textrm{deg},
        \end{equation}
        where $\alpha_{i_k}$ and $\alpha_{i_{k+1}}$ are the inclinations of the well in two consecutive segments along the well trajectory. The segments are assumed to be linear yielding a piece-wise linear trajectory.
        \todo[inline]{ECS: In a piecewise trajectory the gradient/inclination is discontinuous in each node. So maybe you should explain how the inclination is calculated in the nodes? By taking the average inclination on each side of the node? Sergey: The inclination of a line is well defined. We do angle between line segments.}
        \item The inclination is limited to 90 degrees, which is a normal constraint to avoid problems with gravel packing;
        \begin{equation}
            \alpha_{i} \leq 90 \, \textrm{deg}.
        \end{equation}
    \end{enumerate}

\subsection{Individual objectives}
The DSS described in the paper has a simple application programming interface (API) which allows to add new objectives pragmatically. 
It is possible to add objectives that give value for a given point or segment of the well. It is also possible to add objectives as a function of two consecutive segments which, for example, allows to set the cost of bending the well trajectory.
The parameters of the implemented objectives can be changed through the graphical interface.

To simplify the communication in this paper, instead of using conversions to currency, 
we express the value in equivalent number of stands drilled. One unit is equivalent 
to an expected net present value from a well segment positioned in 
a one-meter-thick reservoir layer  with reference properties. 
In this subsection we only list the objectives that were used in the numerical examples (see Figure~\ref{figNewWeights}).
\todo[inline]{ECS: is this a standard from reservoir engineering? (If so, mention it.) Is this valid also for more complex estimations, e.g. of production, water coning, and what-not? If not, a comment could maybe be added?}
\todo[inline]{SEAL: Not a standard. It is something used in the paper. I added words for the paper. Maybe clearer this way.}
\begin{enumerate}
    \item \textbf{Position} in a sand layer. This objective is defined for a well segment and gives value proportional to the thickness $h(x)$ of a sand layer. The value is doubled if the well is positioned in the "sweet spot" for production (in our examples between 0.75 and 2.25 meters from the sand roof). This can be formally written as:
    \begin{align}
        &O_p (x_0, z_0, x_1, z_1) 
        = \dfrac{1}{\delta x_{decision}} \int\limits_{x_0}^{x_1} F_p(h(x)) dx,
        \label{eqPositionObjective}\\
        &F_p(h(x))
        = \left\{
        \begin{array}{ll}
            0, & \textrm{ well is outside reservoir}; \\
            2h(x), & 0.75 \leq z_{roof}(x) \leq 2.25 \textrm{: well in "sweet spot"};  \\
            h(x), & \textrm{otherwise};
        \end{array}
        \right.
        \label{eqThicknessObj}
    \end{align}
    where $(x_0, z_0)$ and $(x_1, z_1)$ are the start and end of a well segment, $z_{roof}(x)$ is the distance from the roof of the reservoir layer to the well,
    \todo{ECS: about $z_{roof}(x)$: $z$ is depth (distance from MSL) and for clarity should not be confused with distance to something else.}
    and $\delta x_{decision}$ is the distance between decision points (reference length of a well stand projected to horizontal axis) equal to 28.56 meters used to normalize the value to the chosen scale.
    The integral in equation \eqref{eqPositionObjective} is evaluated numerically using mid-point rule quadrature.
\item \textbf{Good sand}. This objective gives a value which depends on the sand quality and can be formally written as:
    \begin{align}
        &O_s (x_0, z_0, x_1, z_1) 
        = \dfrac{1}{\delta x_{decision}} \int\limits_{x_0}^{x_1} F_s(x,z(x)) dx 
        \label{eqGoodSandObjective}\\
        &F_s(x,z(x)) 
        = \left\{
        \begin{array}{ll}
           0, & \textrm{ well is outside reservoir}; \\
           7, & \textrm{ well is in the top reservoir layer};  \\
           14, & \textrm{ well is in the bottom reservoir layer}.
        \end{array}
        \right.
    \end{align}
    The value 7 is similar to a reference reservoir thickness from \eqref{eqThicknessObj}. Here, we use a constant value rather than a thickness to highlight that objectives might be expressed differently depending on user preferences.
    \item \textbf{Drilling cost}. The drilling cost objective assigns a cost to drilling the well. When not weighted, the cost of drilling one meter is proportional to one unit, i.e. assumed net present value from a one-meter-long well in a reservoir of one-meter thickness. The objective function can be written as follows:
    \begin{equation}
        O_d (x_0, z_0, x_1, z_1) 
        = \int\limits_{(x_0, z_0)}^{(x_1, z_1)} 
        -0.003 \, ds,
        \label{eqDrillingCost}
    \end{equation}
    where $\int\limits_{(x_0, z_0)}^{(x_1, z_1)}ds$ is an integral along a well segment. Notice that the drilling cost has negative value. With the default scaling (0.003 in \eqref{eqDrillingCost}) the drilling cost of a stand accounts for approximately 8.6\% of the production potential from a one-meter-thick sand layer.
\end{enumerate}
\todo{ECS: explain the value -0.003 as done e.g. for the Position objective}

\subsection{The primary set of objectives}

In most of the examples we are using the Position objective function combined with the drilling cost. The global objective is written as:
\begin{equation}
    O(X|M) = 1.0 \, O_p(X|M) + 1.0 \, O_d(X|M),
    \label{eqPrimaryObjective}
\end{equation}
where objectives $O_p$ and $O_d$ are defined by equations \eqref{eqPositionObjective} and \eqref{eqDrillingCost} respectively.

\subsection{The alternative weighting of objectives used in Section \ref{secInsights}}
\label{secAlternativeObjectives}

In the example from Section \ref{secInsights}, together with the primary objective defined by \eqref{eqPositionObjective} we consider an alternative weighting of the objective functions:
\begin{equation}
    O(X|M) = 0.3 \, O_p(X|M) + 0.7 \, O_s(X|M)  + 1.0 \, O_d(X|M),
    \label{eqAlternativeObjective}
\end{equation}
    where the sub-objectives are defined in equations \eqref{eqPositionObjective}-\eqref{eqDrillingCost}.

\section*{Acknowledgement}
The authors thank Eric Cayeux and Erlend Vefring for insightful suggestions during the paper preparation.

Funding:  This work was supported by the research project 'Geosteering for IOR' (NFR-Petromaks2 project no. 268122) which is funded by the Research Council of Norway, Aker BP, V{\aa}r Energi, Equinor and Baker Hughes Norway.
Aojie
Hong is supported by the 'DIGIRES' project (NFR-Petromaks2 project no. 280473) which is funded by industry partners Aker-BP, DEA, V{\aa}r Energi, Petrobras, Equinor, Lundin and VNG, Neptune Energy as well as the Research Council of Norway.

\section*{Bibliography}
\bibliography{library,manualbib}

\end{document}